\documentclass[10pt,twocolumn,aps,prx,superscriptaddress]{revtex4-2}

\usepackage{amsmath}
\usepackage{graphicx}
\usepackage{color}
\usepackage{upgreek}
\usepackage[linkcolor = blue, citecolor = blue, urlcolor = blue, colorlinks = true]{hyperref}
\usepackage[version=3]{mhchem}
\usepackage{commath,amssymb}
\usepackage{ushort}
\usepackage[dvipsnames]{xcolor}
\usepackage{tikz}
\usepackage{amssymb}
\usepackage{xfrac}
\usepackage{afterpage}
\usepackage{placeins}
\usepackage[]{mdframed}

\newcommand{\redc}[1]{\textbf{\textcolor{red}{#1}}}

\begin{document}

\author{J. de Graaf}
\email{j.degraaf@uu.nl}
\affiliation{Institute for Theoretical Physics, Utrecht University, Princetonplein 5, 3584 CC Utrecht, The Netherlands}

%%% DOCUMENT

\title{The Geometry behind the Glass Transition and Frictional Jamming\protect\\ in Systems of Two-Dimensional Hard Disks}
\date{\today}

\begin{abstract}
\noindent The relation between dynamics and structure in systems of Brownian bidisperse two-dimensional (2D) hard disks with arrested dynamics is examined using numerical simulations. Surprisingly, the suspensions show dynamic arrest at an area fraction of $\phi \approx 0.777$ over a wide range of disk-size ratios. This is in close agreement with the experimental findings of [Lozano~\textit{et al.}, Nat.~Mater.~\textbf{18}, 1118 (2019)]\nocite{lozano2019active} ($\phi \approx 0.776$) for a quasi-2D colloidal suspension of spheres with large-to-small size ratio of approximately $1.4$. Intriguingly, this also matches a jamming transition ($\phi \approx 0.773$ to $0.777$) found experimentally in a 2D bidisperse granular packing of disks for a similar aspect ratio [Naseer~\textit{et al.} \& Vishali~\textit{et al.}, Powders and Grains (2025)]\nocite{naseer2025micromechanics, vishali2025topological}. Adopting a geometric viewpoint allows for the identification of the \textit{floret pentagonal tiling} ($\phi = \sqrt{3}\pi / 7 \approx 0.777343$), which is comprised of congruent (elongated) pentagonal tiles. In analogy to foam and tissue models, it is the presence of this congruent reference state that induces a dynamical transition in the disordered fluid of the disks. That is, the change in dynamics observed both in a uniformly compressed granular medium and a colloidal one at finite temperature are implied to be caused by the same topological mechanism. Extending this reasoning to a \textit{honeycomb lattice} could also explain the experimentally observed onset of caging dynamics ($\phi \approx 0.6$) in a similar bidisperse, quasi-2D system of colloidal spheres [Li~\textit{et al.}, Nature~\textbf{587}, 225 (2020)]\nocite{li2020anatomy}. An outlook is provided on how this concept may be applied to three dimensions (3D): removing one in four particles from a face-centered-cubic arrangement leads to a \textit{square bipyramidal honeycomb} with associated volume fraction $\eta = \pi/(4\sqrt{2}) \approx 0.55536$. The ideas put forward here thus also provide a possible origin for random loose packing of (frictional) hard spheres in 3D and can potentially shed light onto the nature of the glass transition.
\end{abstract}

\maketitle

%%%%%%%%
\section{\label{sec:intro}Detailed Background}
%%%%%%%%

Arrested dynamics in amorphous soft-matter systems has been a topic of significant interest for decades~\cite{sillescu1999heterogeneity, berthier2011dynamical, hunter2012physics, karmakar2015length, manoharan2015colloidal, weeks2017introduction, tanaka2019revealing}. The experimental realization of colloidal glasses made up of (almost) hard spheres~\cite{lindsay1982elastic, pusey1986phase, pusey1987observation}, has given the field a strong impetus to consider the more fundamental question on the nature of the glass transition~\footnote{The formation of a glass is unlike well-known transitions such as freezing or melting. The latter two involve discontinuities in derivatives of the free energy and are therefore readily identified as true transitions. The formation of a glass, however, appears to follow a smooth increase in the viscosity of dozens (if not more) orders of magnitude. This also means that the system typically falls out of equilibrium and the free energy cannot be properly defined.}. In hard-particle systems, the glass transition takes place at a density for which the system exhibits a relaxation-time divergence~\footnote{The more commonly referenced quantity of temperature,~\textit{i.e.}, glass-transition temperature, is not particularly relevant for hard-sphere systems, as it merely scales the unit of time.}, without its structure changing significantly from that of a fluid~\cite{binder2011glassy}. Despite the kinetic nature of this transition and a lack of obvious concurrent structural change, there has been a dedicated effort into obtaining structural signatures in experiment~\cite{widmer2004reproducible, widmer2007study, brambilla2009probing, vivek2017long, hallett2018local, liu2026measurable}, simulation~\cite{schoenholz2016structural, ninarello2017models, tong2018revealing, marin2020tetrahedrality, berthier2023modern, bera2024clustering}, and theory~\cite{parisi2010mean, berthier2011theoretical, janssen2018mode}.

Over the years, structural signals have been found that correlate well with the appearance of slow dynamics at the particle-neighborhood and system level. For example, in three-dimensional (3D) glasses, such indicators are the abundance of icosahedral~\cite{royall2015role} or tetrahedral~\cite{xia2015structural, boattini2020autonomously} neighborhoods in the system. This type of research has recently taken flight with the application of machine learning~\cite{cubuk2015identifying, schoenholz2016structural, paret2020assessing, bapst2020unveiling, richard2020predicting, alkemade2022comparing, alkemade2023improving, oyama2023deep, jung2023predicting, sahu2024structural, sharma2024selecting, jung2025roadmap}. Machine learning appears able to identify subtle differences in local neighborhoods that correlate well with particle dynamics. Yet, these structural insights can be system dependent~\cite{hocky2014correlation}, and those obtained for 3D may not be (readily) transferrable to two-dimensional (2D) systems.

For 2D hard-disk suspensions and their quasi-2D equivalent --- vertically confined hard-sphere systems --- various local characterizations of a particle's neighborhood have been considered. Such analyses typically start by Voronoi tessellating (or equivalently Delauney triangulating) realizations of the system~\cite{song2008phase, zhao2012correlation, morse2014geometric, bormashenko2018characterization, jin2021using, kim2022structural, morse2023local, worlitzer2023pair, du2024rearrangements}. Subsequently, features of the polygonal tiles are analyzed,~\textit{e.g.}, their aspect ratio~\cite{zhang2024anisotropic}, their anisotropy~\cite{rieser2016divergence}, the associated Minkowski tensor coefficients~\cite{schroder2010disordered, kapfer2012jammed, mickel2013shortcomings},~\textit{etc.} These studies have seen success in identifying the appearance of arrested dynamics, as well as the relation between structure and arrest, although frequently the result is subtle and the interpretation not entirely self-evident.

%%%%%%%%%%%
\subsection{\label{sub:experiment}Experimental Inspiration and Open Question}
%%%%%%%%%%%

\begin{figure}[!htb]
\centering
\includegraphics[width=85mm]{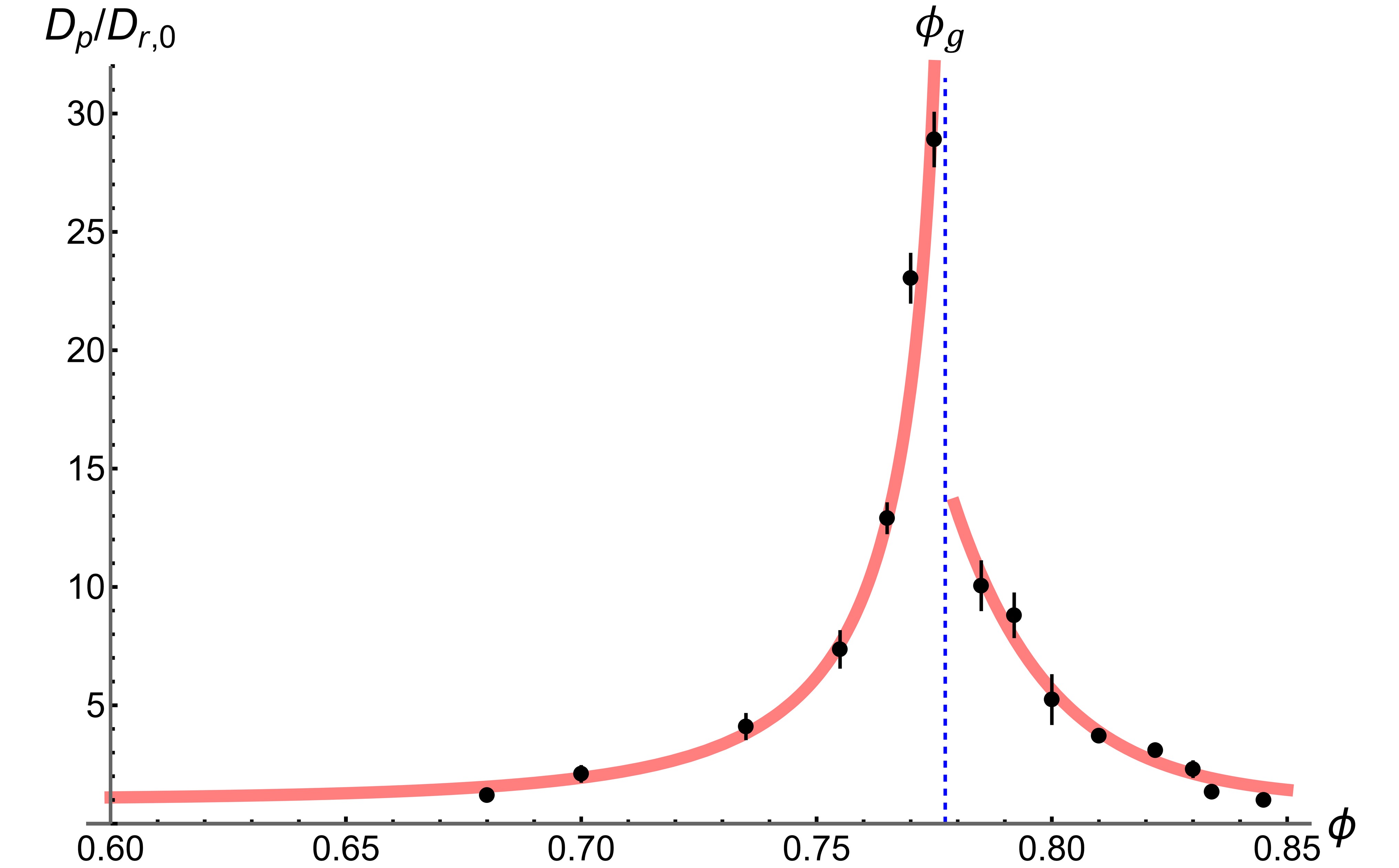}
\caption{\label{fig:lozano}\textbf{Experimental results for the active probe's rotational diffusion.} The effective rotational diffusion coefficient $D_{p}$ normalized by the bare rotational diffusion $D_{0,r}$ of a self-propelled probe as a function of the area fraction $\phi$ of the surrounding bidisperse passive spheres (close to a surface, hence quasi-2D). The black dots show the mean value and the black bars the standard deviation, see Ref.~\cite{lozano2019active}. The two red curves are a stretched-exponential (left) and an exponential (right) fit to the data and serve to guide the eyes. The vertical blue dashed line indicates the value $\phi = \sqrt{3}\pi/7 \approx 0.777$, which will be justified later. Data is reproduced with permission from the authors.}
\end{figure}

Recently, however, an experimental study~\cite{lozano2019active} revealed that self-propelled probe particles give a strong signal close to the glass transition of hard spheres confined to a quasi-2D environment. That is, the orientation of the self-propelled probes was tracked based on the half-coating of their surface, from which the angular mean squared displacement (aMSD) was subsequently obtained. This was found to change upon varying the area fraction $\phi$ of the bidisperse hard-sphere mixture ($R \approx 4.4/6.3 \approx 0.698$; $R$ is the ratio of the small-to-large disk size) that surrounds it. The probe's aMSD  proved linear for the time-differences considered, which allowed for the extraction of the rotational diffusion coefficient $D_{p}$, where the labelling stands for `probe'. The rotational diffusion coefficient proved to be strongly enhanced with respect to its free ($\phi = 0$) value, close to where the translational diffusion in the passive system drops off at $\phi \equiv \phi_{g} \approx 0.776$; even for small self-propulsion speeds, see Fig.~\ref{fig:lozano}. In addition, $D_{p}$ decayed  asymmetrically away from its peak value as a function of $\phi$. In Ref.~\cite{lozano2019active}, this observation was related to the increase in the bulk relaxation time close to their glass transition~\footnote{The glass-transition area fraction was obtained experimentally by studying the behavior of the translational diffusion coefficient of the passive system.} using a theoretical model that incorporated a time-correlated reorientational noise.

It is important to note that in the experiments of Lozano~\textit{et al.}~\cite{lozano2019active}, the $\alpha$ relaxation times obtained from the self-intermediate scattering function follow the Vogel-Fulcher-Tammann (VTF) law~\cite{garca1989theoretical, berthier2011theoretical} exceptionally well. The VFT fit in Ref.~\cite{lozano2019active} indicates a divergence at $\phi~\approx~0.845$, which is close to where random close packing (RCP) for this colloid size ratio sets in, see also the ESI~\cite{ESI}. This calls into question what the nature of the transition at $\phi_{g} \approx 0.776$ is. However, the mean squared displacement (MSD) curves in the experimental system indicate a strong reduction of the passive diffusion coefficient close to $\phi_{g}$. This was related to the system falling out of equilibrium and hence $\phi_{g}$ is where the glass transition is located in Ref.~\cite{lozano2019active}. Here, I will therefore use the term `glass' and `glass transition' thoughout.

Numerical modeling of the probe dynamics in the experimental system~\cite{abaurrea2020autonomously}, performed in my group, showed that an asymmetric peak $D_{p}$ can be obtained using a contact-based friction that induces reorientation. This approach also accounted for the short-time nature of the ballistic component to the aMSD, which occured at times that were not experimentally resolved~\cite{lozano2019active}. However, it proved difficult to place the peak value close to the experimentally observed $\phi_{g} \approx 0.776$. The simulated $D_{p}$ peaks were systematically located closer to the $\phi$ value associated with RCP. In Ref.~\cite{abaurrea2020autonomously}, it was speculated that this may be attributed to a lack of dissipative interactions, such as contact friction or hydrodynamic lubrication, that was not accounted for in our modeling of the passive fluid. In view of the results presented below, the likely orgin is, however, an insufficiently hard-core interaction. A detailed explanation of the experimentally observed transition at $\phi_{g}$ was not provided.

This study was followed up by an experimental and numerical work on a quasi-2D system of colloidal rods~\cite{narinder2022understanding}, which was initiated in the Bechinger group and where my group provided the numerical modeling. Together, it was revealed that there is a $D_{p}$ enhancement for spherical self-propelled probes close to the glass transition of the passive rod fluid, which is well captured using the reorientation model of Ref.~\cite{abaurrea2020autonomously}. Due to the slower dynamics of the rods, we were also able to probe the ballistic enhancement of the aMSD. In addition, considering only the passive (thermal) rod particles proved sufficient to accurately predict the enhancement, meaning that the probes pick up on local features present in the passive system alone. More intriguingly, there was a visually evident structural change in the system close to $\phi_{g}$, namely the breaking up of large rafts of colloidal rods into smaller clusters~\cite{narinder2022understanding}. This suggested that a structural change could have been overlooked in the modeling~\cite{abaurrea2020autonomously} of the experimental system of quasi-2D fluids of hard spheres~\cite{lozano2019active}.

%%%%%%%%%%%
\subsection{\label{sub:granular}Granular Realization and Topological Insight}
%%%%%%%%%%%

For athermal particles,~\textit{i.e.}, ones for which Brownian motion is \textit{not} expected to play a significant role, the jamming transition is considered instead of the glass transition~\cite{majmudar2007jamming, liu2010jamming, behringer2018physics}. This is a transition between flowing and rigid states of amorphous systems. Properties of the networks of contact forces that form in granular materials have been connected with the mechanical response of such systems, including the appearance of the jamming transition~\cite{cates1999jamming, bi2011jamming, sussman2016spatial, berthier2019rigidity}. It is further known that the viscosity $\eta$ of a granular suspension can be fitted using the Krieger-Dougherty relation~\cite{krieger1959mechanism} and that the nature of the forces between particles plays an important role in setting the area (or volume) fraction for which the shear-thickening curve diverges. Experiments~\cite{guy2015towards}, theory~\cite{wyart2014discontinuous}, and simulations~\cite{seto2013discontinuous, mari2014shear, lin2015hydrodynamic, jamali2019alternative, singh2020shear} have shown a branched structure to the response, where the branches are related to the nature of the frictional contacts between particles. These branches control, for example, where RCP (frictionless) and random loose packing (RLP; fully frictional) appear in a hard-particle suspension.

Recent experimental works in 2D~\cite{puckett2013equilibrating, vishali2025topological, naseer2025micromechanics} reveal that the jamming transition in a frictional 2D bidisperse disk suspension ($\phi \approx 0.773$ - $0.777$) lies close to the where Lozano~\textit{et al.}~\cite{lozano2019active} report a glass transition in a colloidal system. In particular, the authors of Ref.~\cite{vishali2025topological} performed a topological analysis on force networks obtained using bidisperse ($R \approx 11/15.4 \approx 0.714$) photoelastic disks subjected to isotropic compression. Their measurements show a topological change around $\phi \approx 0.77$, with the number of homology generators in the persistence diagram of dimension-0 (connected components) saturating at $\phi \approx 0.7773$. This reflects that the isolated force networks merge, which the authors report as predictive of mechanical stability and an early marker of the jamming transition, as also found in earlier work on the topic~\cite{puckett2013equilibrating}. Note that the granular experiments are athermal, while those performed in the Bechinger group~\cite{lozano2019active, narinder2022understanding} are subject to Brownian dynamics. However, the former are performed on an air table and may therefore may have configurations closer to those found in a thermal system than other means of preparing jammed granular packings.

Generally, the jamming and glass transition are considered to have commonalities~\cite{charbonneau2017glass, dinkgreve2018crossover}, but the underlying connection, if present, is not yet fully established. The striking similarity in $\phi$ where the glass transition~\cite{lozano2019active} and jamming transition~\cite{vishali2025topological} are found for two distinct 2D systems begs further examination. The topological nature of the jamming transition obtained in the latter, suggests that there could be a feature shared by both. This is because topological properties enjoy a measure of protection, even in finite-temperature systems. 

%%%%%%%%%%%
\subsection{\label{sub:further}Further Transitions in 2D Thermal Experiments}
%%%%%%%%%%%

One final experimental study needs introduction to help set the stage for this work. In 2020, Li~\textit{et al.}~\cite{li2020anatomy} published results on a quasi-2D confined system of small and large spheres (2.08~$\mu$m and 2.91~$\mu$m, respectively, resulting in $R \approx 0.72 \approx 1/1.4$) prepared at a 0.55:0.45 stoichiometric ratio (close to the 1:1 used in Ref.~\cite{lozano2019active}). The authors of this work perturbed their sample using a focused laser pulse and tracked the particle displacement that this causes. They reported a peak in the associated displacement values around $\phi \approx 0.6$. This peak is collocated with the what the authors refer to as the \textit{onset} of glassy dynamics. That is, Li~\textit{et al.} interpreted the observed peak $\phi$ as being related to the formation of cages and therefore dubbed it an ``onset glass transition''.

The displacement response of the system has similarities to the peak in the rotational diffusion of the self-propelled probe by Lozano~\textit{et al.}~\cite{lozano2019active}. However, given the difference in area fraction and system response, their origin may be different. That is, the MSD data in Ref.~\cite{li2020anatomy} does not reveal a clear drop, as is expected for a glass transition. If such a transition is present in the Li~\textit{et al.} system, it may be that it occurs above $\phi > 0.73$. Lozano~\textit{et al.}~\cite{lozano2019active} did not reveal a convincing probe signal at the values of $\phi$ for which Li~\textit{et al.} find their displacement peak. It may be that the probe's rotational diffusion coefficient starts to rise at around $\phi \approx 0.6$, but it is not possible to verify this based in the data in Ref.~\cite{lozano2019active}. The frictional-reorientation model introduced in Ref.~\cite{abaurrea2020autonomously} shows some enhancement for all values of $\phi < \phi_{g}$, but it may \textit{not} be appropriate for low area fractions.

%%%%%%%%%%%
\subsection{\label{sub:Topgeom}Topological Transitions and Geometry in 2D}
%%%%%%%%%%%

The connection between topology / geometry and response in 2D transitions is elegantly shown in epithelial tissue modeling, and, in particular, in foam-like vertex- and Voronoi-based approaches~\cite{bi2015density, bi2016motility, barton2017active, yang2017correlating, atia2018geometric, yan2019multicellular, krajnc2020solid, damavandi2022universal, claussen2025mean, krommydas2025collective, puggioni2025collective}. When there are $N$ cells, these systems typically possess a Hamiltonian that contains terms of the form
\begin{align}
\label{eq:hamiltonian} \mathcal{H} &= \sum_{i=1}^{N} \left[ \kappa_{A} \left( A_{i} - A_{0} \right)^{2} + \kappa_{P} \left( P_{i} - P_{0} \right)^{2} \right] ,
\end{align}
where $A_{i}$ and $P_{i}$ are the instantaneous area and perimeter length of the $i$-th (Voronoi) cell, respectively. The parameter $A_{0}$ is the target area and $P_{0}$ represents the target perimeter length, which are the rest `sizes' for the harmonic terms in the Hamiltonian. Note that these are common to all cells; heterogeneity in these parameters has only recently been studied~\cite{godolphim2025parameter}.

It is well-established that, in the limit of vanishing cell activity, there is a transition from an amorphous fluid-like state to an amorphous solid-like state when the shape index $p_{0} \equiv P_{0} / \sqrt{A_{0}}$ assumes the critical value $p_{0}^{\ast} \approx 3.81$~\cite{bi2015density, bi2016motility}. The transition value is identical to the $p_{0}$ belonging to a regular pentagon --- a 5-sided polygon for which all edges and angles are equal. The physical interpretation is that this choice of $p_{0}$ leads to frustration, since regular pentagons do not tesselate space, while the foam-like models impose confluence. That is, there is a mismatch between the local requirement on the cells to be pentagonal and the global requirement on the model epithelium to cover all of space.

\begin{figure}[!htb]
\centering
\includegraphics[width=85mm]{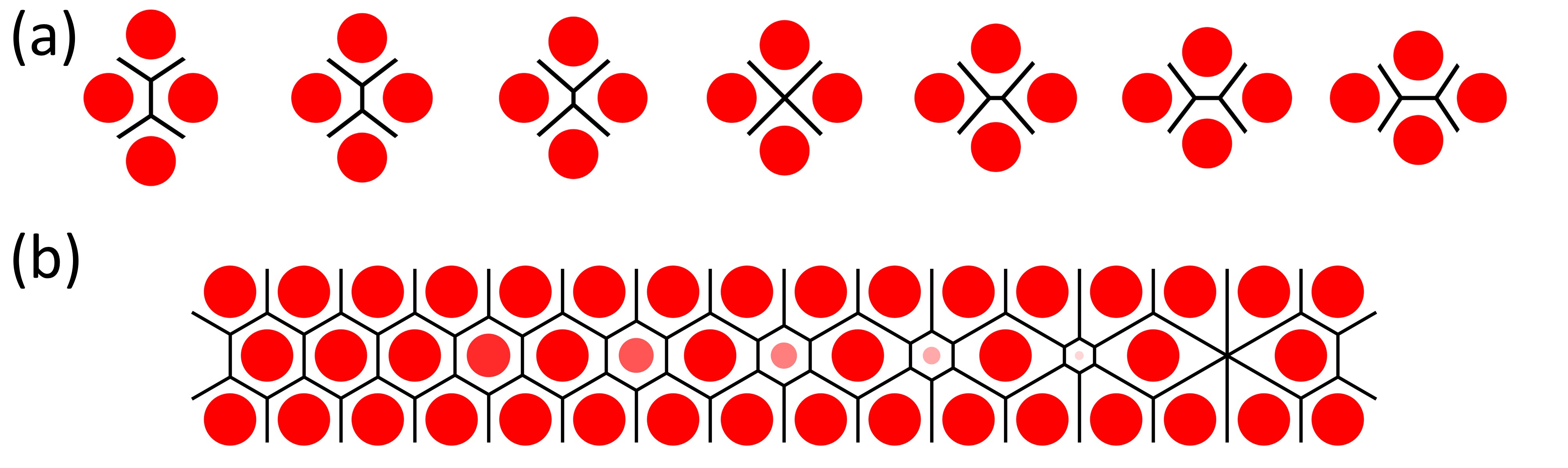}
\caption{\label{fig:transitions}\textbf{Visualization of topological transitions in a 2D system of disks.} (a)~Neighbor exchange of particles (red disks) is a T1 transition, as shown using part of the Voronoi lattice for this configuration (black lines). In the central plot, the particles are equidistantly placed and the transition takes place. (b) In a T2 transition, a particle is removed or added. This is illustrated here by shrinking one of the particles. In the middle row, the disk is smaller for every other particle and made increasingly opaque from left to right, starting from the second particle on the left. The black lines indicate the radical Voronoi diagram for this configuration.}
\end{figure}

Bi~\textit{et al.}~\cite{bi2015density} showed that the onset of arrested dynamics --- here referred to as a jamming transition as the system need not be thermal~\footnote{Technically, this only holds for vertex models in the zero-temperature limit~\cite{sussman2018no}.} --- can be correlated to the appearance of a barrier that hinders topological exchange between neighboring cells. There is also a more obvious transition at the critical value of $p_{0}^{\dagger} = \sqrt{8 \sqrt{3}} \approx 3.72$, which is where the system has hexagonal cells~\cite{staddon2023role}, and it seems likely that this can principle be extended to other space-filling tessellations. Importantly, the topological transition in this system is neighbor exchange, referred to as a T1 transition. There are also T2 transitions, which are associated with the introduction and removal of particles~\cite{weaire1999physics, puggioni2025collective}, see Fig.~\ref{fig:transitions} for a visual representation.

%%%%%%%%%%%
\subsection{\label{sub:goal}Goal of the Paper}
%%%%%%%%%%%

Taking these separate pieces of evidence into consideration, here, I hypothesize that there is a topological feature that underlies both the glass transition in thermal and the frictional jamming transition (at RLP) in granular hard-disk systems in 2D. In addition, I further hypothesize that this topological transition has an elegant geometric interpretation, analogous to the one found in foam and tissue models. Potentially, this hypothesis can be extended into three dimensions (3D), for which I will provide preliminary results in this work.

To investigate the proposed connection, I will circumvent the issues in previously modeling the experimental system using soft potentials~\cite{abaurrea2020autonomously} by studying numerically 2D systems of hard disks using event-driven molecular dynamics (EDMD) simulations~\cite{smallenburg2022efficient}. Here, I will focus on a stoichiometric ratio of 1:1 small to large disks, as in the experiment by Lozano~\textit{et al.}~\cite{lozano2019active}, but I will analyze a far larger range of size ratios $R \in [1/1.7, 1]$ and area fractions $\phi$. This will allow me to determine whether there are lines of near-constant area fraction, for which transitions occur. The presence of these lines means that the particle aspect ratio is not relevant for the transition.

Combined with the experimental evidence that the transition is (potentially) independent of particle size, it becomes feasible to instead examine features of a single component. This simplifies the adoption of a granocentric viewpoint~\cite{corwin2010model, odonovan2013mean, hilgenfeldt2013size}. That is, one where the properties of a neighborhood around a single disk is considered, since this may then be done for a monodisperse system. I will propose a geometric interpretation \textit{via} this route, which points toward a floret pentagonal tiling~\cite{schattschneider1978tiling, alsina2023panoply}. That is, this tiling is a geometric ground state, which underlies the observed dynamic arrest in both thermal and athermal systems. This idea will subsequently be tested further, in order to show the relevance of geometric ground states to the response of 2D fluids of hard particles. The analysis reveals that the result is robust and that a honeycomb-lattice ground state (equilateral-triangle Voronoi tiling) can also impact the dynamics, potentially explaining the observation by Li~\textit{et al.}~\cite{li2020anatomy}.

%%%%%%%%%%%
\subsection{\label{sub:organize}Structure of the Paper}
%%%%%%%%%%%

The remainder of this paper is organized as follows. Section~\ref{sec:method} provides the details necessary to reproduce my results and Section~\ref{sec:character} the methods by which I characterized the structure in the bidisperse hard-disk system. The major features of working with the isoperimetric quotient $q$, which is a variant of the aforementioned shape index $p$ ($q = 4\pi / p^{2}$), are put forward in Section~\ref{sec:struct}. This is followed by an analysis of the dynamics in the system \textit{via} the MSD in Section~\ref{sec:dyna}. Together, these results lead me to propose a state diagram in Section~\ref{sec:diagram}, which is discussed in detail. Next, I correlate structural features to particle dynamics in Section~\ref{sec:relation}, which is supplemented by a literature analysis (Appendix~\ref{sec:compare}). Then, I present my geometric model and understanding in Section~\ref{sec:geound}. The implications of the study are further contextualized in the discussion, where I also give a perspective on an extension to 3D, see Section~\ref{sec:discussion} and Appendix~\ref{sec:3Dper}. Lastly, I provide a short summary of the work in Section~\ref{sec:close}.

%%%%%%%%
\section{\label{sec:method}System and Numerical Approach}
%%%%%%%%

I study structural and kinetic properties of systems of binary hard disks in 2D using the EDMD software developed by Smallenburg~\cite{smallenburg2022efficient}. The algorithms were appropriately modified for use in 2D, see the ESI~\cite{ESI}. This variant of EDMD works in the canonical ($NAT$) ensemble~\footnote{Variants of EDMD that solve for Brownian dynamics are possible~\cite{scala2007event}. Here, the thermalization is introduced by redrawing the velocity of a random selection of particles from the Maxwell-Boltzmann distribution.}. The choice of parameters is such that in the limit of $\phi \downarrow 0$ the (extrapolated) diffusion coefficient of the particles is (typically) given by $D = 1$~\footnote{Note that the collision rate for small $\phi$ is itself small, given the low density of particles. As such, the algorithm does not reproduce diffusion correctly in the limit $\phi \downarrow 0$, which can result in a strong increase of the diffusion coefficient at low $\phi$. This effect is present is in some of the data, but not shown here.}. That is, the reduced time is expressed in terms of the bare diffusion time and I verified that this was the case.

The investigated systems consist of a 1:1 mixture of small and large discs, for which two particles (indexed $i$ and $j$) interact \textit{via} the pair-interaction potential
\begin{align}
U(r) &= \left\{ \begin{array}{cc} 0 & 2r > \sigma_{i} + \sigma_{j} \\ \infty & 2r \le \sigma_{i} + \sigma_{j} \end{array} \right. ,
\end{align}
where $r$ measures the center-to-center distance and $\sigma_{i}$ and $\sigma_{j}$ are the respective particle diameters. I will refer to the diameter of the largest particles as $\sigma$ and write the diameter of the small particles as $R \sigma$, with $R \le 1$ the diameter ratio. Throughout this paper, I will flexibly label graphs by either $R$ or $R^{-1}$, where the latter is often a more precise notation because of parameter choices made in my study. That is, $R^{-1} = 1.7$ has a preference over the notation $R \approx 0.5882352941\cdots$.

In all cases, I simulated $N = $~2,048 particles in a square simulation box with edge length $L$ and periodic boundary conditions~\footnote{I performed a small number of simulations with 10,000 particles, which showed little difference to the results obtained for a smaller number of particles.}. The value of $L$ was varied such that I could simulate area fractions $\phi = (\pi/8) \sigma^{2} \left( 1 + R^{2} \right) N / L^{2}$ between $0$ and $\phi_{m}$. Here, $\phi_{m}$ is the RCP area fraction, which I established for a given $R$ using the approach of Desmond and Weeks~\cite{desmond2009random}, the limitations of which are discussed in Appendix~\ref{sec:compare}. In brief, I computed 250 potential (random) closest packings using the same number of particles $N$ as employed for my production runs. From these I selected the highest value of $\phi$ and thereby obtained an estimate for the RCP area fraction $\phi_{m}$ as a function of $R$, see Section~\ref{sec:diagram} for the result.

I typically considered 100+ non-equidistant points $\phi~\in~[0, \phi_{m}]$ for each value of $R$. For each $R$-$\phi$ combination, I generated 350+ independent realizations of the system by slowly growing the particles until the target value of $\phi$ had been obtained, see the ESI~\cite{ESI}. A dimensionless growth rate of $10^{-3}$ proved sufficient to obtain initial configurations for all $\phi$ even very close to $\phi_{m}$. It should be noted, however, that the time for finding an initial configuration increased substantially as $\phi \uparrow \phi_{m}$, which is expected for the highly constrained packing.

Next, I allowed each configuration to evolve for $10^{4}$ reduced time units without growth, before I started to sample. It will become clear that these $10^{4}$ time units proved sufficient to equilibrate the system for most considered $\phi$-$R$ combinations, despite the slow dynamics. This time only proved insufficient very close to RCP at low $R$ values, see ESI~\cite{ESI}, which are \textit{not} relevant for the conclusions drawn in this study~\footnote{For the purposes of completeness, I should note that in addition to EDMD~\cite{smallenburg2022efficient}, I have attempted to create and equilibrate systems using plain Metropolis, rejection-free~\cite{bernard2009event, michel2014generalized}, and swap variants of the Monte Carlo algorithm~\cite{grigera2001fast, berthier2016equilibrium, ninarello2017models}. However, none of these methods proved to be more efficient than the EDMD algorithm used here. In particular, the use of swap Monte Carlo proved challenging, due to the bidisperse hard-particle nature of the sample at small values of $R$. That is, most moves were rejected beyond a certain value of $\phi$. Though, I should note that by no means this implies that these algorithms are not generally useful to study hard-particle systems with arrested kinetics, for example, as recently evidenced in Ref.~\cite{ghimenti2024irreversible}.}.

%%%%%%%%
\section{\label{sec:character}Characterization}
%%%%%%%%

\begin{figure}[!htb]
\centering
\includegraphics[width=85mm]{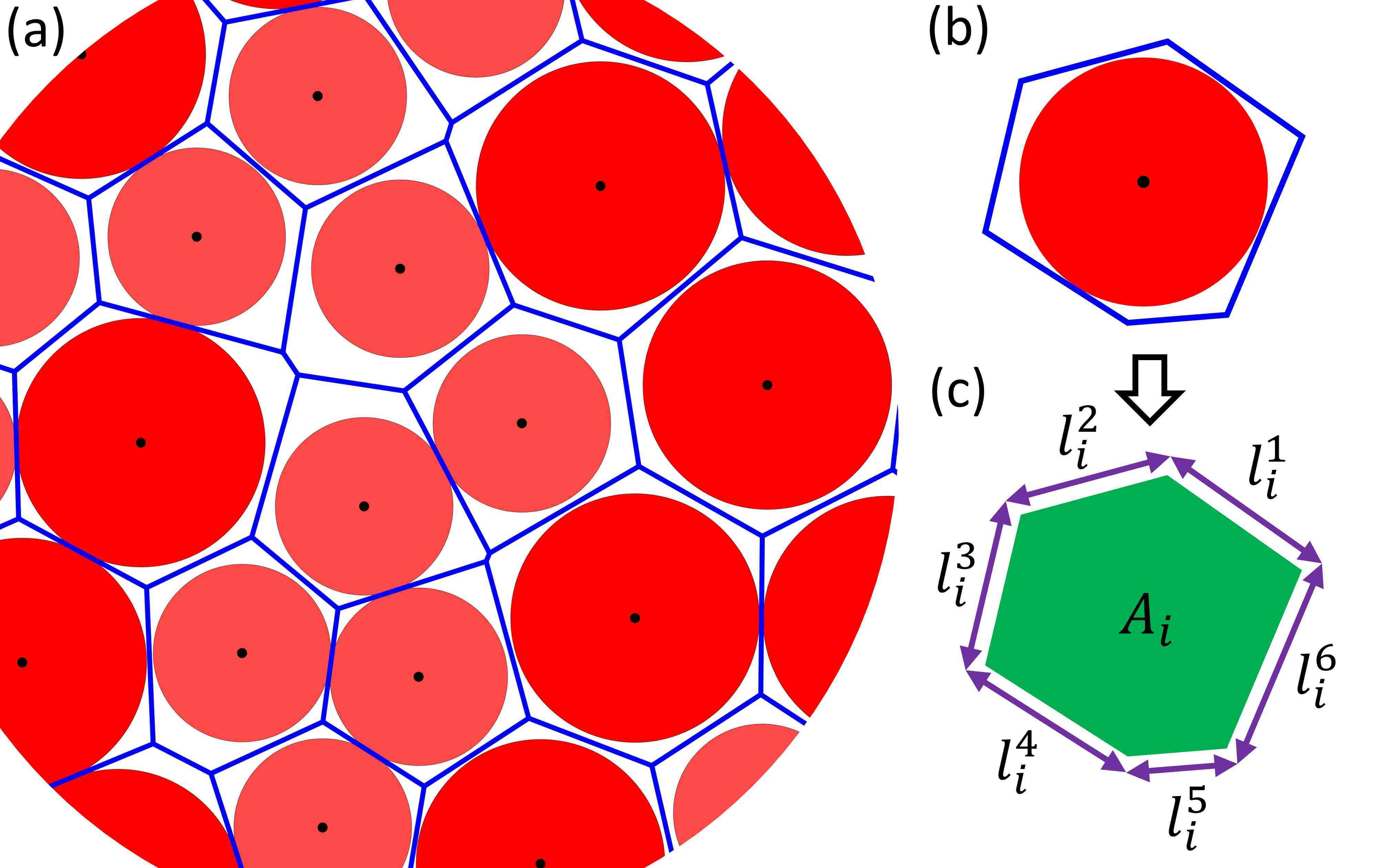}
\caption{\label{fig:voronoi}\textbf{Procedure for computing the isoperimetric quotient using a Voronoi tessellation.} (a) Part of a snapshot of a system showing the instantaneous organization of bidisperse disks with size ratio $R^{-1} = 1.4$ at an area fraction of $\phi = 0.777$. The smaller disks are indicated in light red, while the larger disks are in the darker shade of red. The centers of the disks are indicated using black dots and the blue lines surrounding the disks represent the radical Voronoi diagram for the system. Note that none of the disks overlap though this is hard to resolve by eye. (b) An example of a Voronoi cell for a large particle which has $n = 6$ sides, which is the $i$-th cell in this sample. (c) The area $A_{i}$ of this cell (green) together with the perimeter length $P_{i} = \sum_{i=1}^{n} l_{i}$ (purple arrows) are used to compute the isoperimetric quotient $q_{i} = 4 \pi A_{i} / P_{i}^{2}$.}
\end{figure}

I probed structure by computing the radical Voronoi tessellation (also known as power diagram or Dirichlet cell complex) using the \textbf{\texttt{voro++}} library~\cite{rycroft2006analysis, rycroft2007multiscale}. The radical Voronoi tessellation respects the radius difference between the two sizes of particles in the system. Figure~\ref{fig:voronoi}a shows a result of this analysis for a small part of a simulation volume in for $R^{-1} = 1.4$ at an area fraction of $\phi = 0.777$; close to the experimentally relevant value of the area fraction~\cite{puckett2013equilibrating, lozano2019active, vishali2025topological}.

Using the Voronoi cells obtained in this manner, I computed the isoperimetric quotient
\begin{align}
q &= \frac{4 \pi A}{P^{2}} ,
\end{align}
for each particle, where $A$ is the area of the associated Voronoi neighborhood and $P$ is its perimeter. Note that $q \in (0,1]$, where $q = 0$ corresponds to an extremely needle-like Voronoi neighborhood, while $q = 1$ represents a perfect circle. Recall that for each $R$-$\phi$ combination, there are (typically) $350\times$2,048 $q$ values. This allows me to compute the probability density function (PDF), here denoted $P_{\phi}(q)$, using non-linear binning for $q$ to a high degree of accuracy, as I will present in Section~\ref{sec:struct}.

Importantly, the choice for $q$ rather than $p$ is meaningful, because for specific congruent tilings of the plane, the area fraction of monodisperse disks inscribed in this tiling equals the value of $q$. That is, $\phi = q$ for these special configurations. \textit{The proof is as follows:} 
\begin{mdframed}
Take a tangential polygon --- this has an inscribed circe (diameter $\sigma$) touching all sides --- with $n$ sides of potentially different length. The area of this disk is then $\pi \sigma^{2} /4$ and the polygon's area can be computed by dividing it into $n$ triangles by drawing line segments from the incenter to each vertex. For each side $l_{i}$ the triangle's surface area is $\sigma l_{i}/ 4$. Therefore, the total surface area is $A = (\sigma/4) \sum_{i=1}^{n} l_{i} = \sigma P /4$. Hence, the area fraction is $\phi = (\pi \sigma^{2} /4) /(\sigma P /4)  = \pi \sigma / P$. But, from the area, it follows $\sigma = 4 A / P$ which leads to $\phi = 4 \pi A / P^{2} = q$.
\end{mdframed}
The interpretation useful to this work is that if a constant value of $\phi$ is found, there is an automatic connection to a congruent tiling of the plane. For this tiling all particles must touch, such that the radical Voronoi diagram consists of tangential polygons only.

To improve understanding --- $n = 5$ edges is more transparent than $q \approx 0.865$ --- it proves beneficial to convert $q$ to a \textit{generalized number of edges} $n_{q}$~\cite{nemati2024cellular}, here the subscript ``$q$'' is used to indicate the origin of this number. The isoperimetric quotient of regular polygons with $n$ edges is unique and reads
\begin{align}
\label{eq:nast} q_{\mathrm{r}}(n) &= \frac{ 4\pi A_{\mathrm{r}}(n) }{ P^{2}_{\mathrm{r}}(n) } = \frac{ \pi \sin{(2\pi/n)} }{ 2n \sin^{2}{(\pi/n)} } ,
\end{align}
where $A_{\mathrm{r}}(n)$ and $P_{\mathrm{r}}(n)$ are the area and the perimeter of a regular polygon having $n$ edges and a circumscribing circle with radius $r = 1$. Numerically inverting Eq.~\eqref{eq:nast}, I am able to obtain an $n_{q}$ for any value of $q$, where now non-integer values of the edge number are permissible, such as $n_{q} = 5.2$~\footnote{Note that there is only a single polyhedron that maps to integer values of $n_{q}$, namely the regular one. However, there are many, subtly different polyhedra that give, for example, the value $n_{q} = 5.2$. The point of introducing the generalized number of edges is to see all of these as roughly the same and close to pentagonal.}. Additionally, the definition of $n_{q}$ also allows me to identify ranges in $q$ belonging to triangular, square,~\textit{etc.}~neighborhoods by taking $\lfloor n_{q} \rceil$. For example, all neighborhoods having $\lfloor n_{q} \rceil = 5$ (\textit{i.e.}, $4.5 < n_{q} \leq 5.5$) can be referred to as \textit{pseudo-pentagons}. Integrating the PDF belonging to the pseudo-pentagons provides insight into the fraction of pentagonal neighborhoods in the system, also see Appendix~\ref{sec:frac}.

To complement this Voronoi-based analysis, I also computed the traditional six-fold Steinhardt order parameter $\psi_{6}$, which has the form
\begin{align}
\psi_{6} &=  \left\langle \frac{1}{N} \textstyle \sum_{i = 1}^{N} \left \vert \textstyle \sum_{j = 1}^{N_{n}(i)} \frac{l_{i}^{j}}{P_{i}} \exp \left( 6 \iota \theta_{ij} \right) \right\vert \right\rangle .
\end{align}
Here, $\langle \cdots \rangle$ indicates ensemble averaging, performed over all particles, labelled $i$, and their nearest neighbors $N_{n}(i)$, which were determined using the radical Voronoi construction. The radical Voronoi diagram gives the edge length of the contact between neighbor $i$ and $j$, here written as $l_{i}^{j}$, which is normalized by the total perimeter length $P_{i}$ of the $i$-th Voronoi cell, see also Fig.~\ref{fig:voronoi}. The angle $\theta_{ij}$ between the bond joining the $i$-th and $j$-th particle is with respect to the $x$-axis here, but the choice of this axis has no influence on the result. Finally, $\iota = \sqrt{-1}$ denotes the complex unit and the vertical bars $\abs{\dots}$ indicate taking the norm of the complex vector.

I also computed several ensemble-averaged and time-dependent physical quantities of interest to further support my arguments. These include the pressure $p$, see Appendix~\ref{sec:press}, and two variants of cluster percolation, see Appendix~\ref{sec:perc}. Lastly, I determined the kinetics in the bidisperse hard-disk samples by computing the MSD. I obtained particle displacements for $10^{4}$ unit times starting from the equilibrated samples. This led to 150 MSD independent curves that were averaged over all particles in the system. Subsequently, I determined the mean of the independent MSD data --- the initial configuration varies --- as well as the associated standard deviation. Section~\ref{sec:dyna} provides additional details on the processing of this data, by which I obtained the area fraction $\phi_{a}$ for which the diffusion vanishes.

%%%%%%%%
\section{\label{sec:struct}Neighborhood Structure}
%%%%%%%%

\begin{figure*}[!htb]
\centering
\includegraphics[width=175mm]{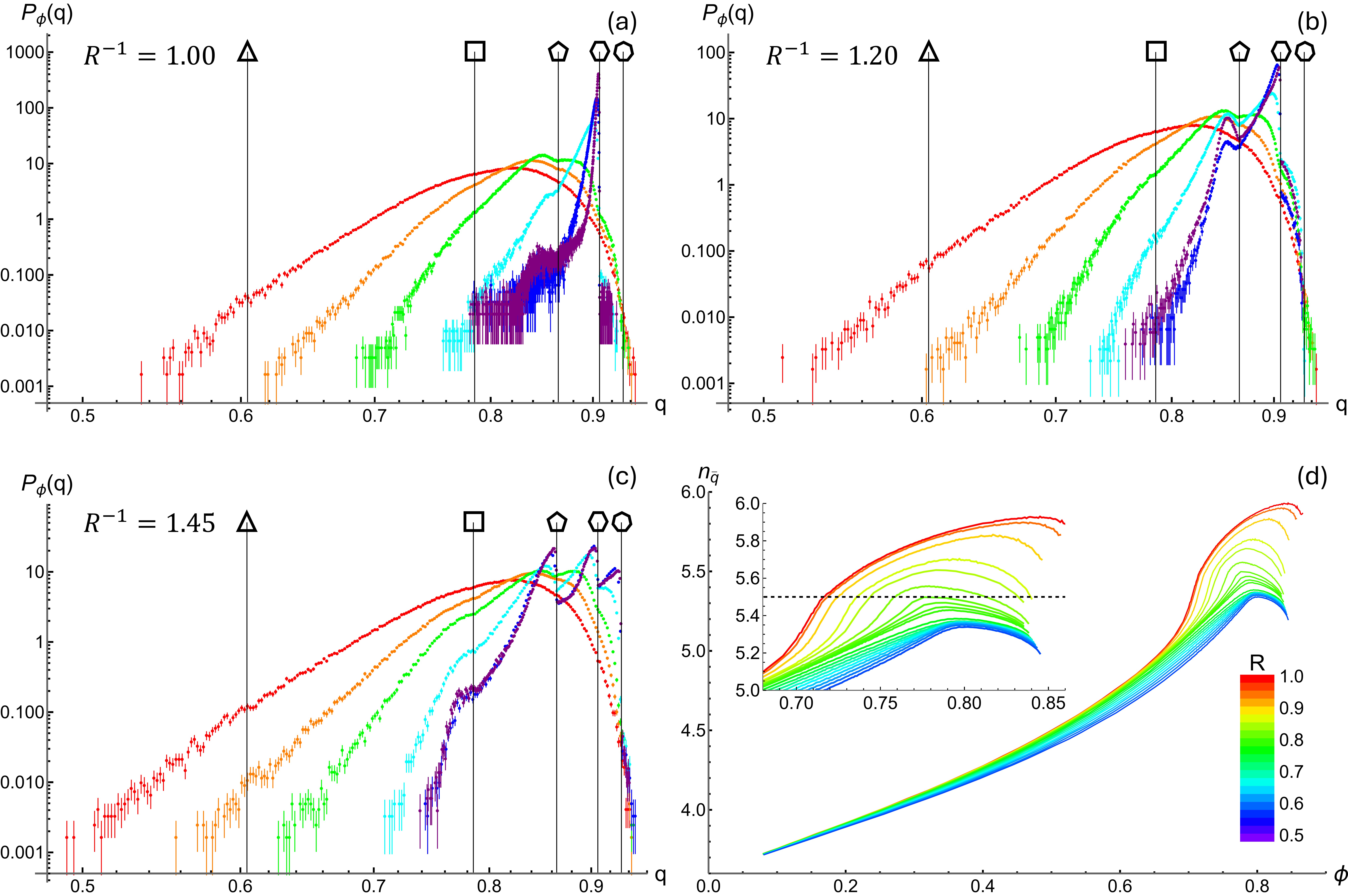}
\caption{\label{fig:pdfs}\textbf{Structural features of the 2D system characterized by the isoperimetric quotient.} (a-c) The probability density function (PDF) $P_{\phi}(q)$ of the isoperimetric quotient $q$ for several values of the area fraction $\phi$ in lin-log representation. The inverse size ratio $R^{-1}$ between the large and small disks in the system is provided in by the label. From red to purple, the area fractions are $\phi = 0.351$, $0.341$, $0.339$ (a-c; red), $0.503$, $0.500$, $0.497$ (a-c; orange), $0.616$, $0.608$, $0.606$ (a-c; green), $0.720$ (cyan), $0.780$ (blue), $0.825$ (purple), respectively. The vertical black lines provide the $q$ values for a regular $n$-gon, with $n = 3,\dots,7$ from left to right; also see visual guide on top. The error bars indicate the standard deviation on the binned value for the interval over which $P_{\phi}(q)$ is sampled, which adheres to counting statistics and is thus always the square root of the number of samples in a bin. (d) The generalized number of edges $n_{\bar{q}}$ belonging to the mean $q$ value --- $\bar{q}$: the first moment of $P_{\phi}(q)$ --- as a function of $\phi$ for all size ratios $R$ considered. From red to purple the value of the size ratio is given by $R^{-1} = 1.00$, $1.05$, $1.10$, $1.15$, $1.17$, $1.20$, $1.22$, $1.23$, $1.24$, $1.25$, $1.30$, $1.35$, $1.40$, $1.45$, $1.50$, $1.55$, $1.60$, $1.65$, and $1.70$, respectively. The inset shows a zoom-in on the range where the systems form crystals,~\textit{i.e.}, above the dashed horizontal line provides the value $n_{\bar{q}} = 5.5$. I do not provide error bars here to help improve the representation. The error is about twice the line width, as can be appreciated from the noise that is present on the red curve in the inset.}
\end{figure*}

Figure~\ref{fig:pdfs}a-c shows a sampling of the isoperimetric-quotient PDFs, $P_{\phi}(q)$, for three values of $R$ and six values of $\phi$, as labelled. Note that irrespective of $R$, the low-$\phi$ diagram is similar. The distribution tends to an established curve as $\phi \downarrow 0$ that was computed for Voronoi diagrams deriving from uniformly distributed points~\cite{zhu2001geometrical, tanemura2003statistical}. This makes sense, as for low $\phi$, properties such as differences in particle size should wash out of the distributions, as also seen in the literature~\cite{morley2020voronoi}.

The results become more interesting when examining intermediate to high values of $\phi$. Note there are discontinuities in the slope of $P_{\phi}(q)$ at specific values of $q$. The $q$ values correspond to those of a regular $n$-gon --- $q_{\mathrm{r}}(n)$ in Eq.~\eqref{eq:nast} --- as is illustrated using the vertical black lines in Fig.~\ref{fig:pdfs}. This is \textit{not} a profound result and can be easily explained as follows. Any polygon with a number of sides less or equal to say $5$ can only contribute to $P_{\phi}(q)$ for $q \le q_{\mathrm{r}}(5)$. However, there are also hexagonal, heptagonal,~\textit{etc.}~neighborhoods that appear pentagonal, but for which $q \gtrsim q_{\mathrm{r}}(5)$. As it is statistically unlikely to have regular pentagons in the sample, this gives rise to a dip in $P_{\phi}(q)$ and an associated slope discontinuity.

Comparing the high-$\phi$ PDFs for the three $R$ values, I observe that the monodisperse ($R=1$) system convincingly crystallizes. The PDF has a single peak which is close to $q_{\mathrm{r}}(6)$, as expected. For lower values of $R$, there remains a sizable fraction of pentagons and there can also be a significant number of heptagons, when $R$ is sufficiently small. I refer to Appendix~\ref{sec:frac} for additional details on the fraction of pseudo-polygons. For systems with $R^{-1} \gtrsim 1.4$, I consistently find that the two peaks in the ranges $(q_{\mathrm{r}}(4),q_{\mathrm{r}}(5)]$ and $(q_{\mathrm{r}}(5),q_{\mathrm{r}}(6)]$ are comparable in height. However, there is a slight abundance of pseudo-pentagons, also see Appendix~\ref{sec:frac}. This means that the disk configurations are not crystalline.

The features of the $P_{\phi}(q)$ will be further analyzed in Section~\ref{sec:relation}, where we will connect them to $\psi_{6}$. However, before turning to dynamics in Section~\ref{sec:dyna}, I will first consider the mean value of $q$ that can be obtained from $P_{\phi}(q)$ by taking the first mode of the distribution. Let this be denoted by $\bar{q}$ and its associated generalized number of edges by $n_{\bar{q}}$. The latter is shown as a function of $\phi$ in Fig.~\ref{fig:pdfs}d for all values of $R$ considered. The curves are generally smooth within the error. There is no apparent gap or discontinuity for those systems that are known to undergo a phase transition. This is because I worked in the canonical ensemble and used a limited number of particles. There appears to be a change in slope close to where the system is expected to crystallize (for $R \lesssim 1$), but I did not investigate this further. The generalized number of edges $n_{\bar{q}}$ is an increasing function of $\phi$ for $\phi \lesssim 0.76$. For greater area fractions, I find a single, well-defined maximum, beyond which $n_{\bar{q}}$ decreases rapidly. I will use this feature, as well as the intersection points with $n_{\bar{q}} = 5.5$, see inset to Fig.~\ref{fig:pdfs}d, in preparing a $R$-$\phi$ state diagram in Section~\ref{sec:diagram}. 

%%%%%%%%
\section{\label{sec:dyna}Particle Dynamics}
%%%%%%%%

\begin{figure}[!htb]
\centering
\includegraphics[width=85mm]{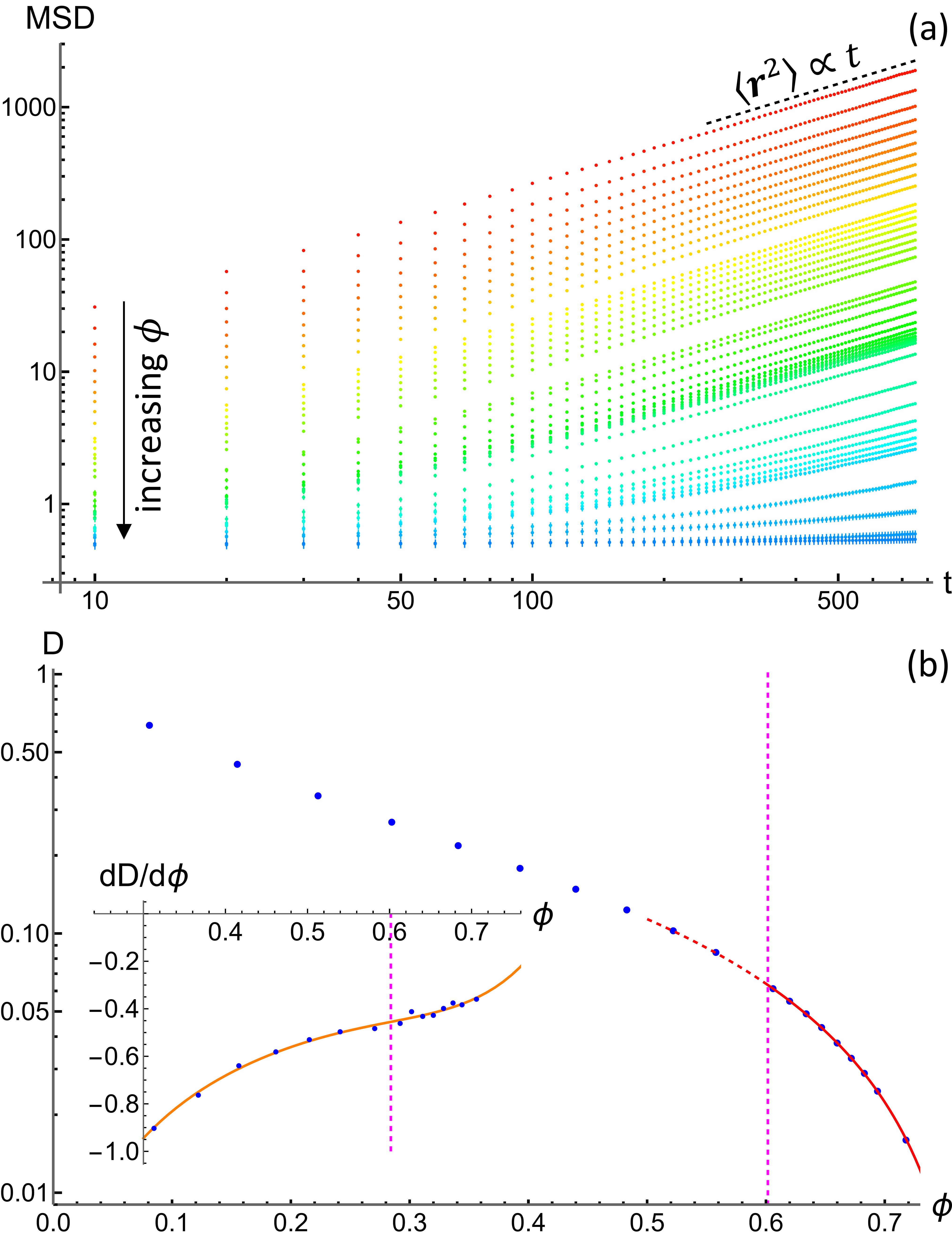}
\caption{\label{fig:diff}\textbf{Analyzing the dynamics in the system.} All the data in this set of graphs was obtained for $R^{-1} = 1.45$. (a)~The reduced mean squared displacement (MSD) as a function of the reduced time $t$. From red to blue the value of the area fraction $\phi$ increases from $0.081$, $0.155$, $0.223$, $0.285$, $0.341$, $0.393$, $0.44$, $0.483$, $0.522$, $0.558$, $0.606$, $0.62$, $0.634$, $0.647$, $0.66$, $0.672$, $0.683$, $0.694$, $0.718$, $0.723$, $0.731$, $0.739$, to $0.744$, respectively. The dots show the average and the error bars the standard error of the mean. The dashed black line shows the linear scaling of the MSD for large times. (b)~Logarithmic plot of the reduced diffusion coefficient $D$ as a function of $\phi$ for the data in (a). The error is typically much smaller than the symbol size. The red curve shows a power-law fit to the data. The vertical dashed vertical magenta line indicates the inflection point to the change in diffusion. The inset shows the numerical derivative $\partial D/\partial \phi$ as a function of $\phi$. The solid orange curve shows a quintic polynomial fit to this data and the magenta line is the same as in the main panel.}
\end{figure}

As explained in Section~\ref{sec:character}, I obtained MSDs for a range of $\phi$ given a specific value of $R$. That is, I computed $\langle \vert \boldsymbol{r}(t) \vert^{2} \rangle$, where $\boldsymbol{r}$ is an individual particle's displacement from (reduced) time $t = 0$ and the angular brackets indicate ensemble averaging. Figure~\ref{fig:diff}a provides examples of such MSD data for $R^{-1} = 1.45$, which is close to the experimental $R$ value of Ref.~\cite{lozano2019active}, as well as a representative sample for the other $R$ considered in this work. Data for which the standard error exceeds 5\% of the mean on the last point is not shown. However, it should be noted that data for values of $\phi > 0.744$ (not shown here) was studied and that within the error in this data, I observe a plateau to the MSD, which is indicative of strong caging.

From the selected curves, I extracted the power-law coefficient $\alpha$ that best described the long-time behavior of the MSD,~\textit{i.e.}, $\langle \vert \boldsymbol{r} \vert^{2} \rangle \propto t^{\alpha}$. The ESI~\cite{ESI} provides full details on the procedure. Whenever this fitting procedure led to an $\alpha \in (0.97,1]$, I used the fit to also obtain the effective, reduced diffusion coefficient $D$. As expected, $D$ decreased as a function of $\phi$, see Fig.~\ref{fig:diff}b. It was found to exhibit power-law decay close to the point where $D$ became vanishingly small. Beyond the associated $\phi$ value, the MSD also typically started to become significantly subdiffusive (again, not shown here).

I fitted the diffusive, power-law decaying part of the MSD using a variable-exponent power law $A \left( \phi_{a} - \phi \right)^b$, where $A$ is a prefactor, $b$ the exponent, and $\phi_{a}$ the area fraction for which the extrapolated dynamics becomes arrested. The value of $b$ was found to increase from $\approx 0.7$ at $R = 1$ to $\approx 1.3$ for $R = 0.8$. Attempts to obtain $b$ for $R < 0.8$ led to data with significant and undesirable level of scatter in $b$ (though the associated $\phi_{a}$ were less scattered). I therefore fitted with a value of $b = 1.3$ below the size ratio $R = 0.8$. Referring forward to Section~\ref{sec:diagram} the idea behind this choice is that once the system no longer crystallizes, the exponent $b$ freezes out.

My approach can raise questions regarding the quality of the fit and precision with which I obtained $\phi_{a}$, as well as its sensitivity to the procedure of preparing the systems. To address the first of these concerns, I approached finding $\phi_{a}$ using structural characteristics as well, see Appendix~\ref{sec:perc}, wherein I use shell percolation~\cite{bug1985interactions}. In brief, there is a small, but measurable feature in the way the shell-percolation point changes with the shell width $\epsilon$ as a function of $\phi$. Note that there is an analogy to this measure and the way contact networks are analyzed in granular packings under compression~\cite{puckett2013equilibrating, vishali2025topological}. Namely, under uniform compression which only slightly perturbs the system, one would expect (on average) contacts to form there where the distances between particles are smallest. When the $\epsilon$-based percolation point changes as a function of $\phi$, this can indicate the presence of nascent load-baring networks, which is what was recently studied using topological analysis in a granular system~\cite{vishali2025topological}.

I found excellent agreement between the $\phi$ for which this feature appears and $\phi_{a}$, in the range where $\phi_{a}$ was observed to be nearly constant. Here, I should emphasize that arrested dynamics, as identified through $\phi_{a}$, is not necessarily indicative of glassy dynamics. It may equally result from the system crystallizing or undergoing another type of phase transition,~\textit{e.g.}, a topological one.

Turning to the second point, it is not clear that the value of $\phi_{a}$ is insensitive to equilibration and preparation. I addressed this issue by performing simulations with equilibration lengths ranging from $10^{3}$ and $10^{5}$ reduced time units (post formation by growth) for several values of $\phi$ and $R$, see the ESI~\cite{ESI}. The changes across this time range to the structure proved to be nominal. This provides confidence in my analysis of both the dynamics and structure around $\phi \approx 0.77$ (and beyond) and contextualizes my earlier statement regarding the quality of the equilibration and sampling.

Having confirmed that the $\phi_{a}$ feature is robust to changes in equilibration time, I consider a final aspect of $D(\phi)$. Using a simple midpoint scheme, I numerically compute the derivative of $D$ as a function of $\phi$, see the inset to Fig.~\ref{fig:diff}b. This data is fitted using a quintic polynomial (orange curve) and from this the inflection point to $\partial D/\partial \phi$ is obtained. I refer to the associated area fraction as $\phi_{i}$ and have indicated it in the inset to Fig.~\ref{fig:diff}b using a vertical dashed magenta line.

In the main panel (Fig.~\ref{fig:diff}b), $\phi_{i}$ appears to be at the edge of where $D(\phi)$ decays slowly. I have suggestively indicated this by using dashing for the red curve to the left of the inflection point, though it should be noted that lin-log representations can be slightly deceptive. The physical interpretation of this point is that $D$ transitions from a change that appears to plateau,~\textit{i.e.}, a tendency of $D$ to decrease linearly with $\phi$, to accelerated decrease. This can be a signal of a new pathway to $D$ reduction becoming available in the fluid. I will discuss the inflection-point data further in Section~\ref{sec:relation}.

I close this section by mentioning that I have also briefly considered the decay present in the self-intermediate scattering function (SISF). It proved challenging to extract $\phi_{a}$ from the SISF data. Referencing the experiments of Lozano~\textit{et al.}~\cite{lozano2019active} this is unsurprising. The relaxation times obtained from their SISF follow the Vogel-Fulcher-Tammann (VTF) law~\cite{garca1989theoretical, berthier2011theoretical} exceptionally well. This can be appreciated from their Fig.~S3a and the re-analysis of their data in the ESI~\cite{ESI}. The area fraction for which the divergence occurs extracted from the VFT fit is $\phi \approx 0.845$, which is close to where I place the $\phi_{m}$ ($\phi$ for RCP)~\footnote{Small departures are expected as the two sizes of colloids used have a little polydispersity and the smaller colloids can be slightly beneath the larger colloids, as their center of mass can be closer to the substrate.}. That is, the experimental data did not reveal a direct relation between properties of the SISF and the glass transition that was experimentally observed. In view of this, I did not pursue further finding $\phi_{a}$ using the SISF, also because there were already two different approaches that were internally consistent. 

%%%%%%%%
\section{\label{sec:diagram}State Diagram}
%%%%%%%%

\begin{figure}[!htb]
\centering
\includegraphics[width=85mm]{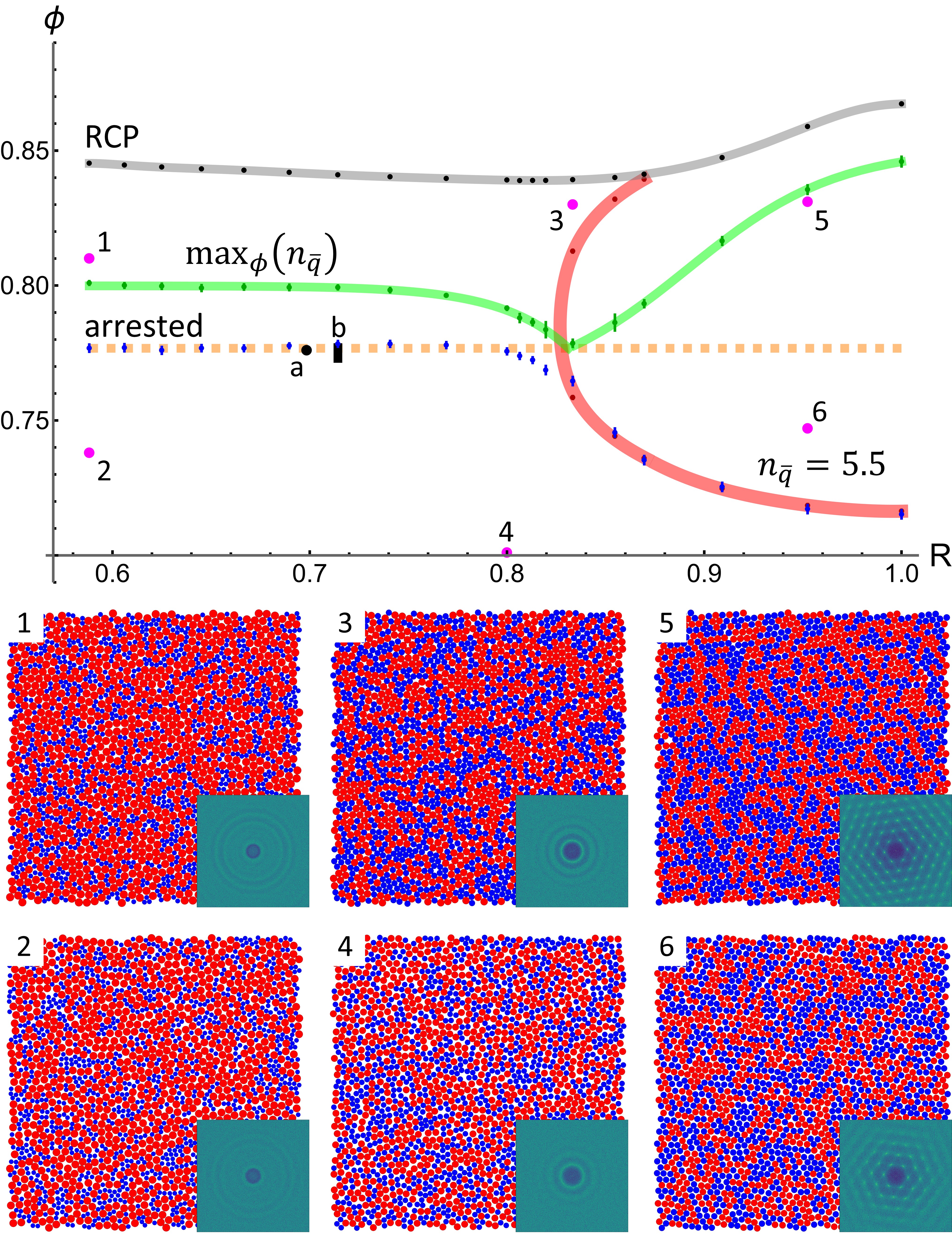}
\caption{\label{fig:diagram}\textbf{State diagram for bidisperse disks in 2D.} The data is represented in a size ratio $R$ to area fraction $\phi$ diagram. Small dots with error bars provide the numerical data and the standard error of the mean, respectively. The solid curves provide guides to the eye for the trends. The black dots (error smaller than symbol size) and light-black curve provide the random-close packing value, labelled ``RCP''. The red dots and light-red curve shows those configurations for which the average generalized number of edges $n_{\bar{q}} = 5.5$. The red curve is thicker than the other ones, as there is more uncertainty as to what its shape is, see main text. The current representation involves a quartic fit near $R \approx 0.83$, see the ESI~\cite{ESI}. The green dots and light-green curves show those systems, for which $\phi$ the value of $n_{\bar{q}}$ is maximized; sigmoidal fits were used here, see the ESI~\cite{ESI}. The blue dots show the points for which the extrapolated dynamics is fully arrested, $\phi_{a}$, as extracted from the MSD. These points partially overlap with the red dots along the bottom red curve. The horizonal, thick-dashed orange line provides the value at which the systems fall out of equilibrium for $R \lesssim 0.8$ and is extrapolated to higher values of $R$. The six numbered magenta dots correspond to the systems, for which representative snapshots are shown at the bottom of the figure, as labelled. In these snapshots the larger disks are indicated in red and the smaller disks ones using blue. The left-bottom corner for each snapshot shows a part of the 2D scattering data. This is provided on a logarithmic color scale, see the ESI~\cite{ESI} for details. The large black dot marked ``a'' indicates the glass-transition value $\phi = 0.776$ for $R \approx 0.698$ as obtained in the colloidal experiment of Ref.~\cite{lozano2019active}. The black bar marked ``b'' indicates the range of values for the jamming point that was obtained in the granular experiments of Refs.~\cite{naseer2025micromechanics, vishali2025topological}, $\phi \approx 0.773$ to $0.777$, with $R \approx 0.714$.}
\end{figure}

From my analysis of the RCP area fraction $\phi_{m}$, the isoperimetric quotient $q$, and the diffusion coefficients $D$, I have put together an $R$-$\phi$ state diagram. This is shown in Fig.~\ref{fig:diagram} and reveals the following trends.

The black points provide the $\phi_{m}$ upper bound for having a (disordered) system that is close packed. Note that for sufficiently high $R \approx 1$, the system is ordered and the notion of `random' in RCP is abused, as snapshot~5 clearly reveals crystalline packing. The algorithm of Desmond and Weeks~\cite{desmond2009random} does not check for disorder, only for close packing, but I will use the term RCP throughout, nonetheless. The light-black curve is a high-order polynomial fit to the data points and serves to guide the eye. The width of this (and other) curve(s) in the diagram corresponds roughly to the maximum deviation from the mean (standard error of the mean) in the data.

I correlate the presence of crystallinity throughout the system with $n_{\bar{q}} > 5.5$. For $R \gtrsim 0.83$ there are two intersection points for $n_{\bar{q}} = 5.5$,  see Fig.~\ref{fig:pdfs}d and its inset, which leads to two distinct branches (red points) that must meet up in a single `critical' point. In Appendix~\ref{sec:press}, I show that $n_{\bar{q}} \ge 5.5$ is where the bond-orientational order parameter predicts a hexagonal crystal. Snapshots~5 and~6 also reveal that the system is crystalline inside of this envelope, while snapshots~3 and~4 reveal disorder. Note that shapshot~3 has several crystalline areas in it, though the overall structure is clearly not that of an ordered solid, as also evidenced by the scattering profile provided in the insets to the snapshots. Analyzing the sample using $n_{\bar{q}}$ is thus meaningful.

It proved difficult to find the `critical' point even for my fine-meshed $R$ analysis; I therefore fitted the data points closest to $R \approx 0.83$ using a polynomial of the form $R = A \phi + B \phi^2 + C \phi^4$ with $A$, $B$, and $C$ fit coefficients. This together with another quartic polynomial fit of $\phi$ in terms of $R$ on the data closer to $R = 1$, see the ESI~\cite{ESI}, leads to the red curve (envelope to the crystalline region) presented in Fig.~\ref{fig:diagram}. The curve for this envelope is thicker than that used for the other guides to the eye in the diagram, as there is more uncertainty on it, especially close to the `critical point' in terms of $R$. Here, I use this wording, because the value of $R \approx 0.83$ appears to delimit the crystalline region in this manner. Note that because there is low diffusivity close to $\phi_{a}$, the analogy to an equilibrium critical point of a temperature-density phase diagram is tenuous. Nonetheless, I will argue for a topological transition se
tting in from this point.

\clearpage

Appendix~\ref{sec:press} also shows results for the pressure as a function of $\phi$. I use this to demonstrate that there is a reasonable agreement between the crystalline-phase coexistence density, as measured using the van-der-Waals loop, and the lower $\phi$-branch resulting from the $n_{\bar{q}}~=~5.5$ criterion. There is also satisfactory agreement with $\phi$ coexistence values previously reported in the literature for bidisperse hard-disk systems~\cite{huerta2012towards}. All of this together leads me to conclude that crystallization region is accurately located in the diagram. Finally, obtaining the hexatic transition was \textit{not} a goal in this study and considering the system sizes required to accurately determine this value~\cite{engel2013hard, qi2014two} is \textit{not} valuable or needed.

Figure~\ref{fig:diagram} also shows the value of $\phi$, for which $n_{\bar{q}}$ assumes its maximum (green data). Note that the data clearly has two trends to it, which were individually fitted using a sigmoidal curve. This leads to the solid, light-green curve that also serves to guide the eye. Above $R \approx 0.83$, I find that $n_{\bar{q}}$ shows a descending trend with decreasing $R$, while below this value of $R$, $n_{\bar{q}}$ increases with decreasing $R$. The transition between these two regimes coincides with the closing of the $n_{\bar{q}} = 5.5$ (red) envelope within the error. The two trends are caused by the same effect. Above the $\phi$ value associated with $\max_{\phi} \! \left( n_{\bar{q}} \right)$, the average neighborhood becomes increasingly anisotropic to accommodate the target area fraction. This could be caused by a global distortion of neighborhoods or by the incorporation of defects. I will provide additional evidence in Section~\ref{sec:relation}.

The trends displayed by the green points may be explained as follows. For the systems that crystallize --- within the $n_{\bar{q}} = 5.5$ envelope --- the lower the value of $R$, the more freedom there is to distort neighborhoods, due to the presence of smaller particles. Thus, the value of $\phi$ for which distortion occurs, decreases with decreasing $R$. Outside of the envelope,~\textit{i.e.}, for $R \lesssim 0.83$, the system is better characterized by disorder, which is occasionally interspersed with crystalline patches, see also snapshots~1,~2, and~4. Therefore, the smaller the value of $R$, the greater the area fraction must be to begin to see distortions in the neighborhoods of the small particles, which contribute 50\% to the value of $n_{\bar{q}}$.

Lastly, I examined the point that follows from the intersection of the green curves, which is located at $\phi \approx 0.777$ with $R^{-1} \approx 1.20$. The area fraction belonging to this intersection value is indicated over the entire $R$ range using the horizonal, thick-dashed orange line. I will motivate this extended range more carefully in the upcoming Section~\ref{sec:relation}. This value of $\phi$ is extremely close to what was reported for the experimental systems of Lozano~\textit{et al.}~\cite{lozano2019active} and Refs.~\cite{naseer2025micromechanics, vishali2025topological}, see point ``a'' and line ``b'' in the state diagram~\ref{fig:diagram}, respectively.

Interestingly, the values of $\phi$ for which the (extrapolated) diffusion coefficient vanishes, $\phi_{a}$, correspond well to $\phi = 0.777$ for sufficiently low $R$. That is, $\phi_{a}$ matches the constant dashed orange curve well for $R \lesssim 0.8$ or equivalently $R^{-1} \gtrsim 1.25$. For $R \gtrsim 0.833$ ($R^{-1} \lesssim 1.2$), I further observe that $\phi_{a}$ accurately tracks the lower part of the red $n_{\bar{q}} = 5.5$ envelope. This makes sense, as diffusion in a crystal is mediated by interstitial mechanisms~\cite{bennett1971studies}, which means that it is generally much slower than diffusion in a fluid. My fitting procedure for $\phi_{a}$ picks up on this difference in dynamics. The more surprising result is the near constancy of $\phi_{a}$ over a large range in $R$, which is what I will turn to next.

%%%%%%%%
\section{\label{sec:relation}Structural Signals of\protect\\ Changes in the Dynamics}
%%%%%%%%

The fact that there is a relation between reduced dynamics and the $n_{\bar{q}} = 5.5$ data for $R \gtrsim 0.83$, inspired me to look for similar indicators of $\phi_{a}$ in the isoperimetric-quotient data for $R \lesssim 0.8$, where arrested dynamics appears to set in at nearly constant $\phi$. Here, I therefore consider the properties of the $q$ distribution and comment on the inflection points obtained in Section~\ref{sec:dyna}.

%%%%%%%%%%%
\subsection{\label{sub:pentagonal}Large-Particle Pentagonal Neighborhoods}
%%%%%%%%%%%

\begin{figure*}[!htb]
\centering
\includegraphics[width=175mm]{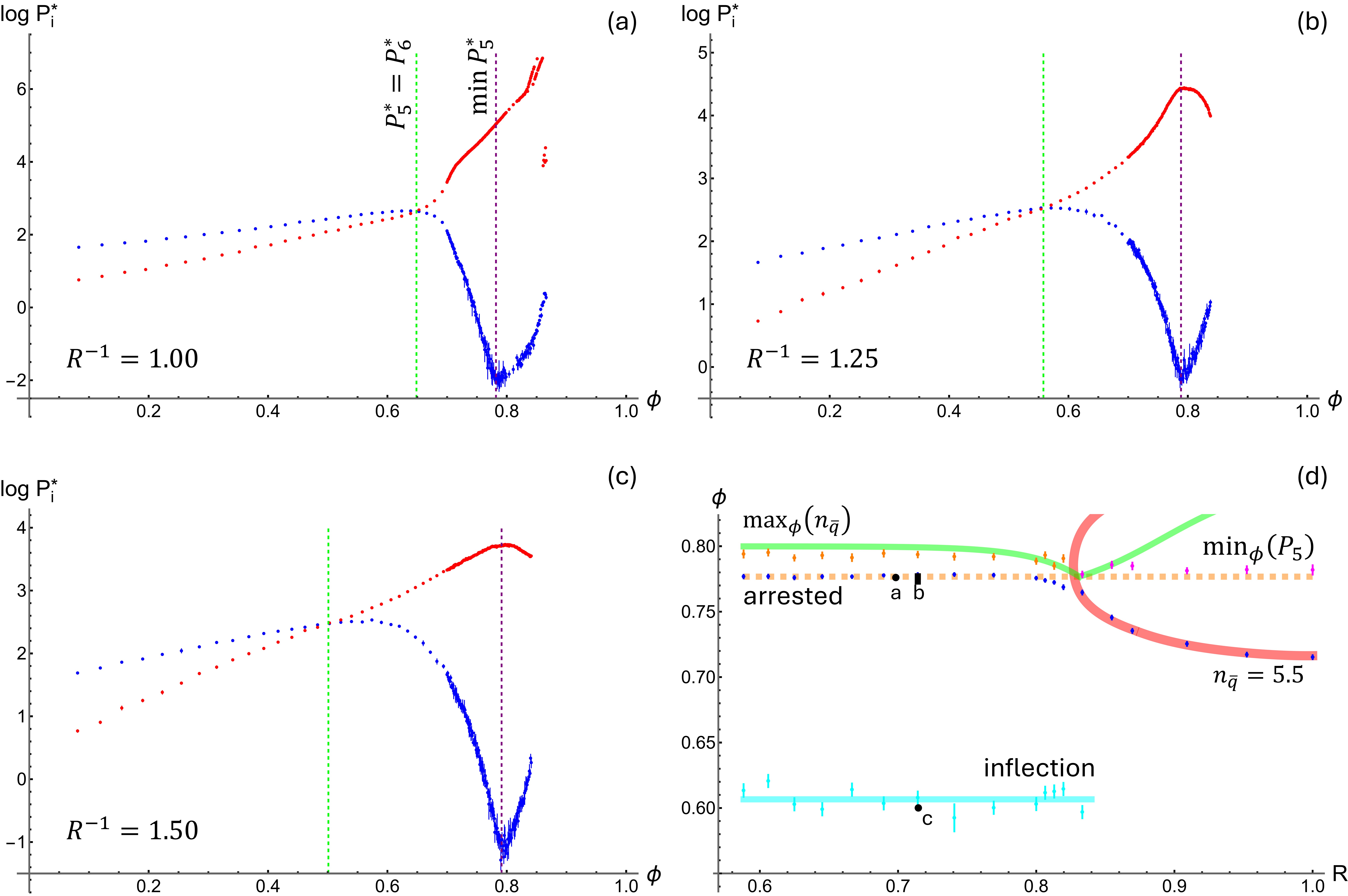}
\caption{\label{fig:pdfpr}\textbf{Identifying the arrested dynamics \textit{via} the isoperimetric-quotient distribution.} (a-c)~The (natural) logarithm of the isoperimetric-quotient PDF for \textit{large} particles only, evaluated at its domain-restricted mode value $P^{\ast}_{i}$, as a function of the area fraction $\phi$. The main text details the procedure by which this data was obtained. Blue points show the value for the submode restricted to the domain $q_{\mathrm{r}}(4) < q \le q_{\mathrm{r}}(5)$ ($P^{\ast}_{5}$) and red points are obtained from data restricted to $q_{\mathrm{r}}(5) < q \le q_{\mathrm{r}}(6)$ ($P^{\ast}_{6}$), respectively. The vertical dashed purple line indicates the local minimum; in this representation this is the location of the cusp in $P^{\ast}_{5}$. The vertical dashed green line indicates the $\phi$ for which the red and blue curves cross each other for the first time. In all cases, the value of the size ratio is labelled in the bottom-left corner of the panel. The error bars indicate the standard error of the mean, though in panel (a) there is a clear set of outliers, caused by limited equilibriation and subsequent issues in the fitting. These were ignored in my analysis. (d)~Reproduction of part of the $R$-$\phi$ state diagram of Fig.~\ref{fig:diagram}. The orange and magenta data shows $\phi$ associated with the minimum in the large-particle $P^{\ast}_{5}$ and the cyan data indicates the inflection point in the diffusion data, see Section~\ref{sec:dyna}. The black point labelled ``a'' corresponds to the glass transition found in Ref.~\cite{lozano2019active} and the black bar labelled ``b'' indicates the range for the jamming transition from Refs.~\cite{naseer2025micromechanics, vishali2025topological}. The black point labelled ``c'' represents the ``onset glass transition'' as reported in Ref.~\cite{li2020anatomy}. All thick, light-colored curves serve as guides to the eye.}
\end{figure*}

First, consider the \textit{large-particle-only} distribution,~\textit{i.e.}, the PDF that is obtained by only considering the $q$ values of the large-particle neighborhoods (the radical Voronoi diagram is \textit{not} altered); this will be justified shortly. This large-particle-only distribution is cut up into subdomains running from $q_{\mathrm{r}}(i-1) < q \le q_{\mathrm{r}}(i)$ with $i=4$, $5$,~\textit{etc.} On each of these subdomains, I establish the (sub)mode of the distribution --- the value of $\phi$ for which the highest (local), large-particle-only PDF is found. Examining the peak values amplifies subtle changes in the underlying distributions, though one should be careful not to overinterpret the outcome. The goal of this section is solely to show that extrapolating $\phi \approx 0.777$ to $R = 1$ is justifiable, even if the effects that are picked up on are small. Here, I specifically examined pentagonal and hexagonal neighborhoods, as these are most prevalent in the bidisperse disk system at high values of $\phi$, also see Appendix~\ref{sec:frac}. The value of the PDF associated with this submode, denoted $P^{\ast}_{i}$, is shown in Fig.~\ref{fig:pdfpr}a-c as a function of $\phi$ for the two domains $i=5$ (blue; square to pentagon) and $i = 6$ (red; pentagon to hexagon), respectively.

Note that $P^{\ast}_{5}$ has a clear minimum, see Fig.~\ref{fig:pdfpr}a-c. Taking the natural logarithm of $P^{\ast}_{5}$ gives this minimum the appearance of a cusp. Both sides of this cusp were subsequently fitted using polynomials and the intersection point computed~\footnote{Note that $P^{\ast}_{5}$ is a positive number always, hence the cusp is not a sign of a number going through zero, as is often found in logarithmic representations where the absolute of a value is plotted.}. The associated value of $\phi$, say $\phi^{\ast}$, is shown in Fig.~\ref{fig:pdfpr}d, where the error bar gives an indication of the accuracy of the fit. The value of $\phi^{\ast}$ for $R > 0.83$ is indicated in magenta and for $R < 0.83$ in orange, respectively. The orange data appears to follow the $\max_{\phi} \! \left( n_{\bar{q}}\right)$ curve, albeit at a slightly lower value of $\phi$, which hints at a correlation betweent the two measurements. For $R > 0.83$ the data appears nearly constant and closer to the $\phi = 0.777$ line. I used this indicator to justify extending the $\phi = 0.777$ line to $R = 1$.

I also examined the distribution $P_{\phi}(q)$, which derives from \textit{all} particle neighborhoods, and the PDF that results by considering only those neighborhoods that belong to \textit{small} particles. In both distributions, the effect for the square-to-pentagon subdomain was clearly present for values of $R \uparrow 1$, but for $R \lesssim 0.83$ there were only small hints of a feature in the distribution, which proved difficult to fit, see the ESI~\cite{ESI}. This further justifies my choice for examining only the large particles. Moreover, large particles, especially for $R \ll 1$, should be the least mobile and, thus, arrested dynamics should originate with these particles. This could explain why I see the strongest signal originating from this fraction.

The data shows that the $q$ associated with $P^{\ast}_{5}$ shifts ever closer to $q_{\mathrm{r}}(5)$ as $\phi$ is increased,~\textit{i.e.}, the neighborhoods become increasingly pentagonal. The peak-value first increases, then decreases rapidly, see the ESI for additional details~\cite{ESI}. In all cases, the decrease sets in as the large-particle neighborhoods tend toward hexagonal,~\textit{i.e.}, beyond the large-particle crossover $P^{\ast}_{5} = P^{\ast}_{6}$. This is when the large particles start to form hexagonal-like clusters to allow the system to further densify. Note that whenever the system has a sufficiently large size asymmetry, the presence of the small particles interferes with the clustering and $P^{\ast}_{6}$ reduces as $\phi$ is increased further, as local neighborhoods distort~\cite{rieser2016divergence}. Following the vertical purple dashed lines in Fig.~\ref{fig:pdfpr}a-c, there is a strong correlation between this downward trend in $P^{\ast}_{6}$ and the presence of the minimum in $P^{\ast}_{5}$ (for sufficiently small $R$). However, the maximum in $P^{\ast}_{6}$ proved more difficult to fit, so that I could not ascertain the degree of correlation.

Intriguingly, the minima in $P^{\ast}_{5}$ are also present for $R \gtrsim 0.83$, as indicated by the magenta data in Fig.~\ref{fig:pdfpr}d. It appears that the mean value of $\phi^{\ast}$ on this domain is slightly lower than that for smaller $R$ (orange data). That is, $\phi^{\ast}$ lies closer to the orange dashed line. This shows, to the best of my understanding, a structural indicator of the system changing within the crystalline phase. However, it should be noted that the effect is difficult to obtain and quite subtle.

%%%%%%%%%%%
\subsection{\label{sub:inflection}Inflection Point for the Diffusivity}
%%%%%%%%%%%

Figure~\ref{fig:pdfpr}d also shows the inflection point data (cyan) that was obtained from the measurement of the diffusion coefficient, see Section~\ref{sec:dyna}. This could only be accurately established for $R \lesssim 0.83$, hence it is not shown over the entire range. There is a substantial error in the data, because it is a difficult point to extract. Note, however, that within this error, the associated value of $\phi$ is effectively constant. This is similar to the way $\phi_{a}$ behaves on this part of the $R$ domain. The guide to the eye in Fig.~\ref{fig:pdfpr}d is a fitted constant with value $\phi \approx 0.607$ to all the data points. This value lies close to where Li~\textit{et al.}~\cite{li2020anatomy} locate their onset glass transition or equivalently the appearance of caging in the fluid, see the black point labeled ``c'' in Fig.~\ref{fig:pdfpr}d. This is added here under the assumption that the small change in stoichiometry in Ref.~\cite{li2020anatomy} (0.55:0.45 rather than 1:1) does not affect the system significantly. Thus, the appearance of the inflection point can be given a caging interpretation, meaning that a new pathway for $D$ reduction becomes available. Note that the caging transition could not be readily obtained from the data that was available in Lozano~\textit{et al.}~\cite{lozano2019active}, as was discussed earlier in the introduction.

%%%%%%%%
\section{\label{sec:geound}Geometric Understanding}
%%%%%%%%

Turning to the literature on foams and tissues~\cite{bi2015density, bi2016motility, barton2017active, yang2017correlating, atia2018geometric, yan2019multicellular, krajnc2020solid, damavandi2022universal, claussen2025mean, krommydas2025collective, puggioni2025collective}, I consider it possible (and plausible) that a geometric feature underlies the observations of dynamic arrest in the disk systems. This would be analogous to the way the pentagonal shape controls the onset of jamming in model tissues. To identify this feature in the disk systems, I adopt a granocentric view~\cite{corwin2010model, odonovan2013mean, hilgenfeldt2013size}. This choice is also informed by the apparent connection to frictional jamming and RLP, for which the method was originally developed. Following this approach, I studied a single disk surrounded by a single neighbor shell of disks.

It proved difficult to explain features of the state diagram as a function of $R$ using this approach on a bidisperse system. That is, it is straightforward to show that a small particle can be stabilized by being surrounded by 5 larger particles (all in contact), leading to a pentagonal neighborhood. At contact, this occurs for $R \approx 0.701302$; the ESI~\cite{ESI} provides the analytic expression. Large particles can have a perfectly heptagonal neighborhood, whenever they are surrounded by 7 small particles (all in contact). At contact, this occurs for $R \approx 0.766422$, also see the ESI~\cite{ESI}. Neither of these values corresponds particularly well with the `critical point' to the crystalline region (as a function of $R$), which occurs for $R \approx 0.83$. It should be clear that the presence of such neighborhoods is a rarity. Thus, I would not expect the appearance of arrest to be captured well by such an analysis. However, the granocentric approach did prove beneficial in making the connection to Voronoi-based foam and tissue models, when applied to a monodisperse system.

The above evidence points to the significance of the area fraction $\phi \approx 0.777$ (and potentially $\approx 0.607$). It is reasonable to assume that the aspect ratio $R$ is not significant in setting the transition criterion, given that $\phi_{a}$ is nearly constant over a wide range of $R$. Note that here it is important that this holds for amorphous packings, and is thus different from the appearance of constant $\phi$ lines in (quasi-)crystalline packings~\cite{fayen2023self}. Using the extrapolation to $R \uparrow 1$ as I argued for above, it becomes possible to examine the properties of a monodisperse system only for clues to the origin of this value of $\phi$. Additionally, the experimental results for a wide range of length scales~\cite{puckett2013equilibrating, lozano2019active, vishali2025topological}, appear to indicate that the particle size $\sigma$ and temperature $T$ do \textit{not} control the transition. To help support the former insight, I make the connection to the literature on experiments and simulation of frictional systems in Appendix~\ref{sec:compare}.

%%%%%%%%%%%
\subsection{\label{sub:ftp}The Floret Pentagonal Tiling}
%%%%%%%%%%%

\begin{figure}[!htb]
\centering
\includegraphics[width=85mm]{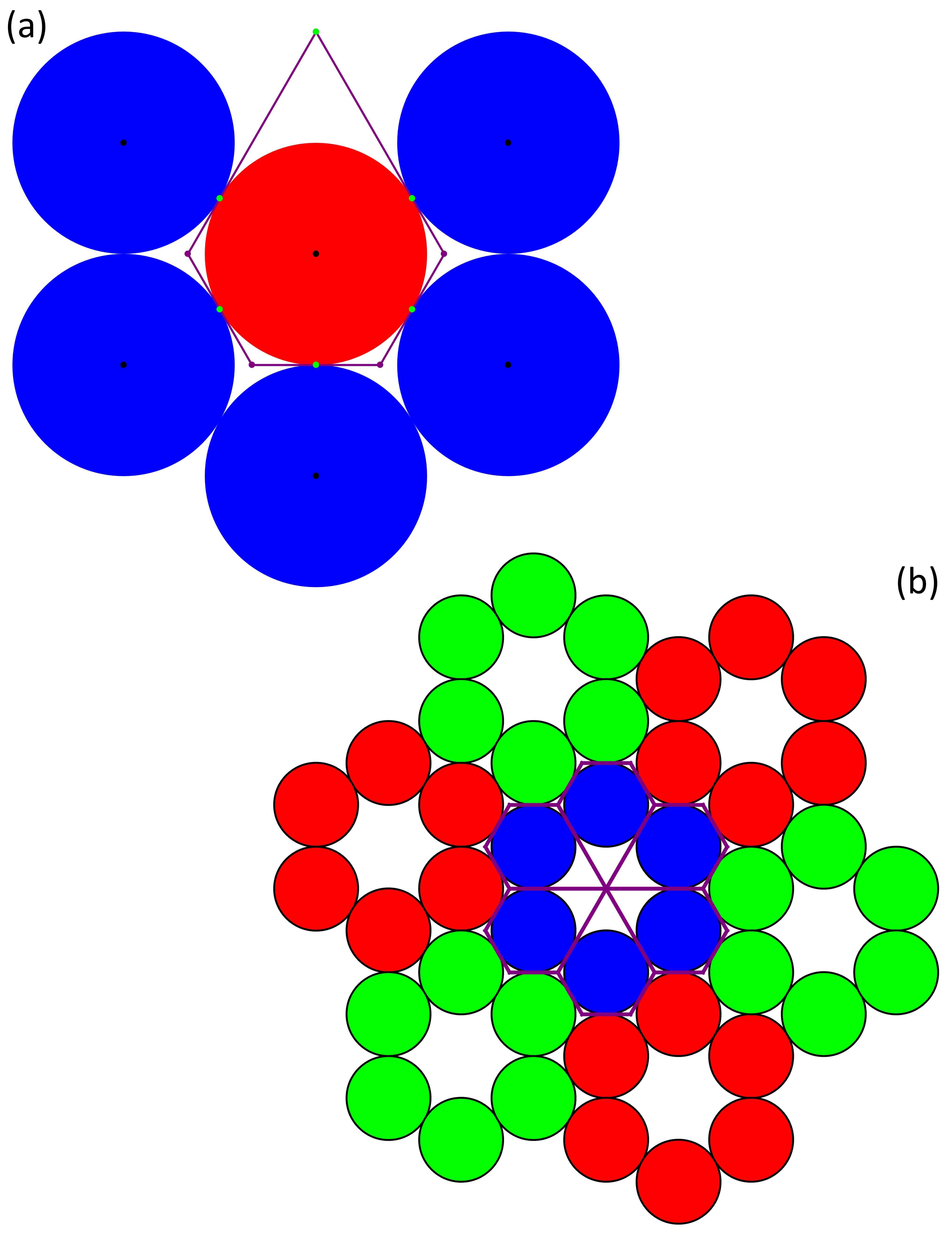}
\caption{\label{fig:geometry}\textbf{Geometric ground state that can be related to the glass transition.} (a) The Voronoi cage (purple) around a single particle (red) surrounded by 5 neighboring disks (blue). These particles are in contact (green points). The black dots indicate the centers of the respective disks. (b) Organizing the particles into wheel-like neighborhoods, wherein all the Voronoi cells have five vertices. The central wheel (blue) shows the 6 pentagonal Voronoi cells (purple). The six surrounding wheels show that these can be stacked into a crystalline structure. The use of alternating red and green coloring is meant to help distinguish the wheels, but otherwise carries no physical significance.}
\end{figure}

Considering instead hexagonal neighborhoods of equal-sized disks, reveals the following. Figure~\ref{fig:geometry}a shows a single, pentagonal neighborhood of a central disk surrounded by five neighbors. This configuration can be seen as a limiting state of moving a sixth neigboring disk upward until it is no longer a Voronoi neighbor with the central particle. It is easy to demonstrate, see the ESI~\cite{ESI}, that the area fraction associated with this central particle is $\phi = \sqrt{3}\pi / 7 \approx 0.777343$, which is remarkably close to the value of $\phi$ for which dynamical arrest was observed in the event-driven simulations.

Interference with hexagonal crystallization --- leading to amorphous states with dynamic arrest --- should ideally be shared equally among all particles. The pentagonal neighborhood identified in Fig.~\ref{fig:geometry}a can be assembled into a regular structure that is effectively a close-packed hexagonal crystal with one particle out of every 7 missing, see Fig.~\ref{fig:geometry}b. For each collection of 6 particles surrounding a void, henceforth referred to as ``wheel'', every particle is in contact with 5 neighbors. Thus, every associated Voronoi cell has exactly 5 vertices. This also means that each particle is completely immobilized, independent of whether it is frictional or non-frictional. The area fraction of this configuration is $\phi = \sqrt{3}\pi$ and it is identical to the isoperimetric ratio $q$.

The structure in Fig.~\ref{fig:geometry}b is known from the field of pentagonal tilings,~\textit{i.e.}, it is a monohedral, convex, pentagonal tiling of type 6~\cite{schattschneider1978tiling, alsina2023panoply}. It is also often referred to as a floret pentagonal tiling (FPT). I will refer to the particle stacking into ordered wheels as an FPT structure in recognition of this connection. 

%%%%%%%%%%%
\subsection{\label{sub:ground}Geometric Ground States}
%%%%%%%%%%%

The FPT structure can also be seen as the result of T2 removals of one in seven of the particles in a hexagonal close packed arrangement, see Fig.~\ref{fig:transitions}b. In particular, one where the removals lead to a regular sublattice of voids. Interestingly, starting from a crystalline solid (at infinite pressure), this is the first (lower) area fraction achieved by T2 removals, for which both the voids are arranged regularly, and all neighborhoods are identical. Equivalently, all Voronoi tiles are congruent and their edges are by construction tangent to the inscribed sphere. This is important, since --- as mentioned previously --- for a monodisperse system where the disks are inscribed to a congruent (tangential) tiling, the equivalence $\phi = q$ holds. That is, the area fraction obtained from the MSD and shell-percolation criterion is an indicator of the relevance of the Voronoi cell belonging to this tiling.

A second such configuration exists, where two out of three particles are removed, with the remaining ones forming a regular honeycomb lattice. This corresponds to a Voronoi tessellation where each cell is congruent and an equilateral triangle. The associated area fraction $\phi = \pi / (3 \sqrt{3}) \approx 0.6046$, very close to our fitted constant value of $\phi_{i} \approx 0.607$ for the inflection point to $\partial D / \partial \phi$. Note that this is removed from where the monodisperse system has liquid-hexatic coexistence ($\phi \approx 0.699$ for the liquid and $\phi \approx 0.716$ for the hexatic phase coexistence area fractions), as well as the continuous hexatic-solid transition that is located at $\phi \approx 0.724$~\cite{strandburg1988two, gasser2009crystallization, engel2013hard, qi2014two, kapfer2015two}. Therefore, it is reasonable to say that the transition related to the inflection point is \textit{not} related to the hexatic phase transition.

For the FPT-based transition there may be a relation to the hexatic phase transition. Above, I related the FPT structure to T2 removals, which can form pairs of pentagons and heptagons in a hexatic lattice. This kind of pair formation is also reported in defect unbinding for the hexatic phase transition~\cite{engel2013hard, kapfer2015two, krommydas2025collective}. In addition, the diffusivity data $\phi_{a}$ runs from $\approx 0.7$ to the value $\phi \approx 0.777$ in a(n apparently) smooth fashion, see Fig.~\ref{fig:diagram}, given the quality of the data. Thus, the FPT-based transition, which I claim to underlie the appearance of arrested dynamics, may be a limiting case of the topological defect unbinding. This connection will need to be analysed further in future work.

Both the FPT and honeycomb correspondence suggest that there is merit to identifying regular geometric ground states that result from T2 removal of particles from the close-packed structure and possess congruent Voronoi tiles. It is possible to form other structures with congruent Voronoi tessellations, but these do not arise naturally from T2 transitions of the close-packed crystalline structure and are therefore not considered here. I revisit this possibility again in 3D in the Discussion section, but there too the most promising connection follows by considering T2 particle removals starting with the closest packed structure.

Note that at zero temperature (infinite pressure) the structure in Fig.~\ref{fig:geometry}b is a crystalline solid and mechanically rigid. This begs the question of how these structures influence clearly disordered states of bidisperse particles. I hypothesize that transforming a geometric ground state can be accomplished by bringing together two bulk FPT states and introducing a few defects into the structure. These defects allow for freedom of motion and a systematic reshaping of the structure to something that looks otherwise like an amorphous system. Alternatively, the Mermin-Wagner theorem --- forbidding long-ranged order in 2D --- could provide the necessary fluctuations to destroy the FPT order and leave an amorphous state at \textit{finite} pressure (and temperature). The influence of the state could then be explained by the topological nature of the transitions required to rearrange particles, neighbor exchanges involve T1 transitions \textit{not} T2 transitions. It is well known that topological properties are invariant under diffeomorphisms, as are other invariants~\cite{hermann2021noether}. The mechanism for the particle system would be similar to the way the isoperimetric target value of the Hamiltonian~\eqref{eq:hamiltonian} controls the dynamics of a model foam or tissue, as $p_{0}$ is not influenced by neighbor exchanges.

For the honeycomb lattice, there is no intrinsic rigidity (save for fully frictional particles) and smoothly distorting this configuration into that of an (arrested) amorphous fluid phase is easier to understand. That is, I expect that the value of $\phi_{i}$ for a bidisperse system can be explained by similarly bringing together two perfect honeycomb lattices with different particle sizes and allowing these to rearrange \textit{via} thermal motion into a disordered fluid-like state. Unlike for the FPT state where $P_{5}^{\ast}$ showed a clear cusp, I found no hints of features in the local structure  at $\phi \approx 0.607$ that could justify taking the limit $R \uparrow 1$ for this near-constant value of $\phi_{i}$. This makes the geometric undestanding more tenuous in for this value of $\phi$. I will address these issues next by performing additional computations.

%%%%%%%%%%%
\subsection{\label{sub:validation}Validation}
%%%%%%%%%%%
 
Formalizing mathematically the above understanding is not trivial. For the foam and active tissue models, the connection to (soft glassy) rheology is made through the trap model~\cite{kollmannsberger2009active, bi2016motility, nguyen2025origin}. But given the hard nature of the interaction in the bidisperse disk system investigated here, this may not be readily transferable. Mechanical topological features were only recently discussed in the context of hyperform materials~\cite{hong2024topological}. However, it is possible to give an additional piece of information to support my viewpoint of geometric ground states.

If such a state is indeed responsible for the dynamic arrest, then the area fraction $\phi \approx 0.777$ should also appear from other compositions of the system. Turning to Appendix~\ref{sec:compare}, this may explain why point ``h'' in Fig.~\ref{fig:compare}, which was obtained for another stoichiometric ratio (1:2)~\cite{donev2007configurational}, falls within the range reported by Refs.~\cite{naseer2025micromechanics, vishali2025topological}. This connection can be made stronger by performing an additional numerical study on a sufficiently different system in terms of radii and composition.

\begin{figure}[!htb]
\centering
\includegraphics[width=85mm]{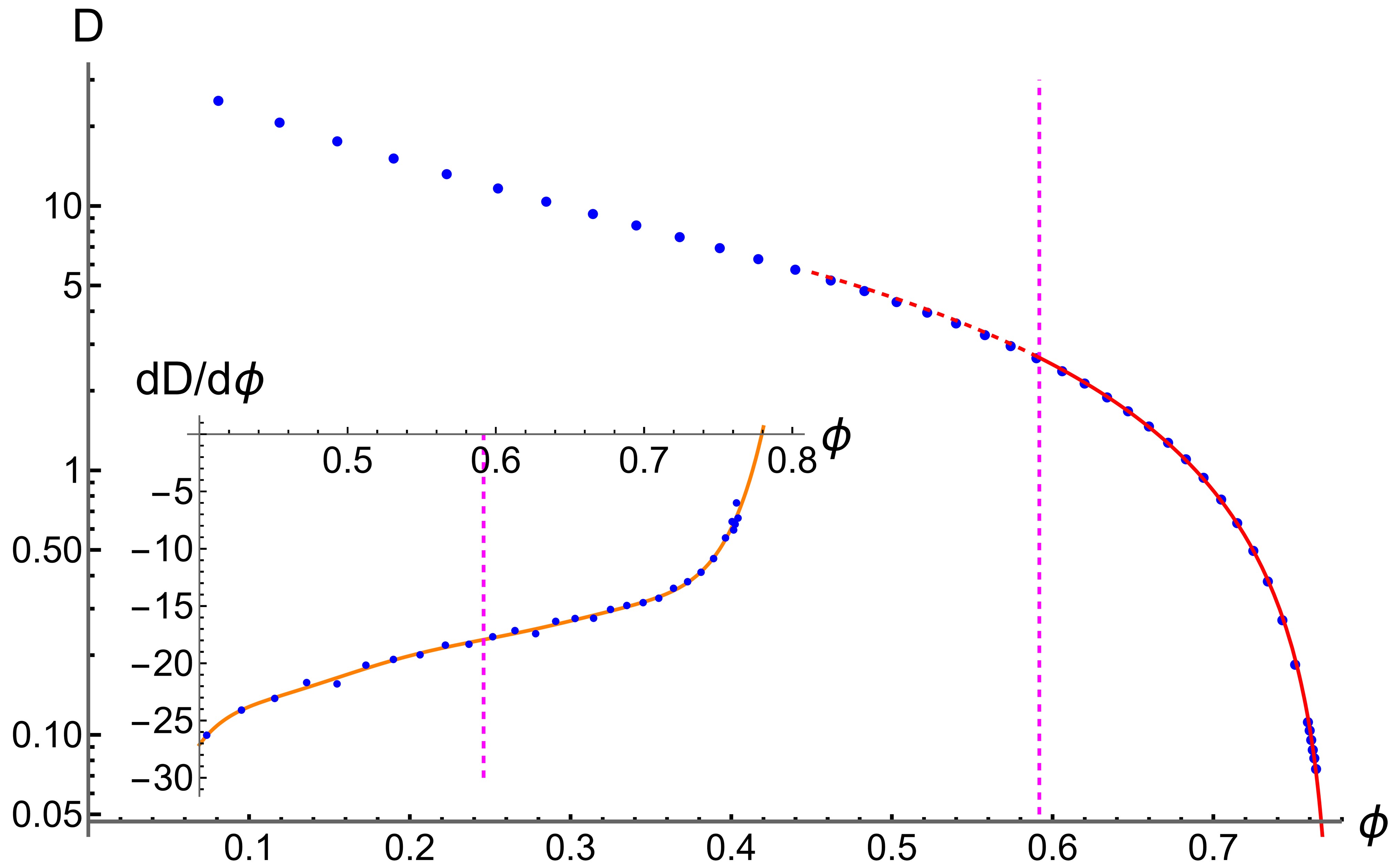}
\caption{\label{fig:valid}\textbf{Validating the geometric ground state} Logarithmic plot of the reduced diffusion coefficient $D$ as a function of $\phi$ for a trinary with particle sizes $\sigma$, $0.8 \sigma$, and $0.6 \sigma$ in 1:1:1 stoichiometric ratio. The error is typically much smaller than the symbol size. The red curve shows a power-law fit to the data. The vertical dashed vertical magenta line indicates the inflection point to the change in diffusion. The inset shows the numerical derivative $\partial D/\partial \phi$ as a function of $\phi$. The solid orange curve shows a septic polynomial fit to this data and the magenta line is the same as in the main panel.}
\end{figure}

Figure~\ref{fig:valid} shows the diffusion coefficient obtained for a system of 2,400 hard disks with sizes $\sigma$, $0.8 \sigma$, and $0.6 \sigma$ in a 1:1:1 stoichiometric ratio. Data was processed in a manner analogous to that for our bidisperse system, although I equilibrated the system for longer, namely by $10^5$ diffusion times. I found the extrapolated dynamical-arrest point to be located at $\phi_{a} = 0.775 \pm 0.004$ and the inflection point at $\phi_{i} = 0.59 \pm 0.04$. Note that it remained difficult to localize the inflection point accurately. However, it is clear that the area fraction belonging to the floret pentagonal tiling was recovered, which I further confirmed by analyzing features of the shell-percolation transition, see the ESI~\cite{ESI}.

%%%%%%%%
\section{\label{sec:discussion}Discussion}
%%%%%%%%

Given the above results, there are several points worth critiquing. It is well known that systems with $R \approx 1$ tend to locally crystallize~\cite{donev2007configurational}, rather than locally form an FPT structure~\footnote{Technically the Mermin-Wagner theorem forbids this structure in 2D at finite temperature and pressure.}. That is, the disks can make their density more homogeneous by locally arranging into a hexagonal crystalline arrangement rather than an FPT. Thermodynamically, the state in Fig.~\ref{fig:geometry}b is also unfavored for $T >0$, because there is only a single realization, for which all Voronoi cells are identical and pentagonal. When assuming a hexagonal crystal configuration at $\phi \approx 0.777$, the particles can vibrate about their ideal hexagonal-lattice position, raising both the configurational and vibrational entropy of the system.

Thus, it is difficult to imagine that a geometric ground state, as introduced here, would be able to influence (crystalline) configurations in a thermal or Brownian suspension. However, the numerical evidence presented here, combined with 2D results from other sources~\cite{donev2007configurational, silbert2010jamming, zhang2024anisotropic, hinrichsen1990random, bideau1986geometrical, bolton2024ideal}, see Appendix~\ref{sec:compare}, in addition to the experimental work by Lozano~\textit{et al.}~\cite{lozano2019active} and Li~\textit{et al.}~\cite{li2020anatomy} for Brownian systems, and that of Naseer~\textit{et al.}~\cite{naseer2025micromechanics}, Vishali~\textit{et al.}~\cite{vishali2025topological}, and Puckett and Daniels~\cite{puckett2013equilibrating} for granular disks, strongly suggests the following. The departure away from monodispersity (and potentially from frictionless interactions) allows for the geometric ground states that I report to compete with the driving forces that induce crystallization. This would explain why the indications for the presence of such states is so weak for nearly monodisperse systems, although there is circumstantial evidence in the simulation and theory literature~\cite{hinrichsen1990random, bideau1986geometrical}.

It would be interesting to see to what extent experimental evidence along the lines of the work carried out by Vishali~\textit{et al.}~\cite{vishali2025topological} or frictional simulations like those of Refs.~\cite{silbert2010jamming, singh2020shear} are able to validate this viewpoint, as well as help identify topological features near $\phi \approx 0.777$. However, a force-network analysis and shear-based rheology constitute separate studies and are therefore left for future work. Narinder~\textit{et al.}~\cite{narinder2022understanding} reported a correlation between the structure of a passive rod fluid and the rotational dynamics of an active particle that interacted with this fluid. A contact-frequency argument was made to explain this connection. In Appendix~\ref{sec:fluctuations}, an analysis of the time-dependence of the change in neighborhoods is provided, also see the ESI~\cite{ESI}, as a setup to such an analysis applied to the disk system studied here. This reveals that there are indeed rapid dynamics in the form of neighbor exchanges, which peaks (weakly) at a value close to $\phi_{g}$. This is a piece of evidence in support of the conceptual framework introduced in my group to explain the rotational diffusion enhancement for a self-propelled probe particle~\cite{abaurrea2020autonomously, narinder2022understanding}. Referencing Appendix~\ref{sec:fluctuations}, the peak appears more pronounced for a monodisperse sample. Examining the properties of (quasi-)2D arrangements further using self-propelled probes or laser-type perturbations~\cite{li2020anatomy} might reveal more information, however, this too constitutes a separate project.

The concept of a geometric reference state is by no means a new one. Yet, it is typically applied locally, for example, when examining the dynamics in glasses~\cite{tong2018revealing, tanaka2019revealing, marin2020tetrahedrality, li2020anatomy, kim2022structural, sharma2024selecting, yu2024universal}. What has not been previously reported, to the best of my understanding, is that seemingly disordered (2D hard-disk) particle fluids can become `topologically entangled' with ordered, geometric ground states. Nonetheless, this is not a peculiar way of thinking about 2D systems, when referencing the literature on foam and tissue models~\cite{bi2015density, bi2016motility, barton2017active, yang2017correlating, atia2018geometric, yan2019multicellular, krajnc2020solid, damavandi2022universal, claussen2025mean, krommydas2025collective, puggioni2025collective}. It seems plausible that elements of 2D models based on geometric harmonic potentials be transferred in an effective manner to particles that interact via short-ranged (highly repulsive) potentials. This then suggests a possible common origin behind observations of a glass and jamming transition in 2D hard disks and model tissues, where the former is controlled by the FPT and the latter by an overlapping arrangement of regular pentagons. A potential route toward tackling this problem is to find a transformation that smoothly takes one proposed geometric ground state into another. I see projections of higher-dimensional regular lattices on a 2D subspace \textit{via} a cut-and-project type operation~\cite{duneau1985quasiperiodic, kalugin19856D, elser1986diffraction} as a viable route of exploration in this direction. Speculating, the angle of the plane would have to be varied between rational and irrational to achieve the desired transformation.

Lastly, the notion of a geometric ground state may be transferred to 3D systems. In Appendix~\ref{sec:3Dper}, I analyze several candidate structures. In foam-like tissue modeling, there is evidence that a geometric frustration like the one in 2D can occur~\cite{merkel2018geometrically}. In 3D, the random loose packing (RLP) of spheres is reported for a volume fraction $\eta \approx 0.555$ in various studies~\cite{onoda1990random, silbert2010jamming, guy2015towards, anzivino2023estimating}. This value can be obtained by removing one in every four particles from a face-centered cubic (fcc) arrangement, leading to $\eta = \pi / (4 \sqrt{2}) \approx 0.55536$. The associated space-filling honeycomb (the 3D equivalent of a planar tiling) consists of congruent square bipyramids and is referred to as a square bipyramidal honeycomb or oblate octahedrille, see the ESI~\cite{ESI} for additional information. However, it should be noted that various values of $\eta$ for (frictional) jamming are available, ranging from $\eta \approx 0.54$~\cite{song2008phase, silbert2010jamming} to $\eta \approx 0.64$~\cite{berryman1983random}.

It seems likely that the details of the surface interactions and preparation procedure matter. However, it may also be that there are competing regular honeycombs that interfere with obtaining a uniform value across experiments, be those in a lab or using a computer. For example, I obtain a volume fraction of $\eta = 8 \pi /(27\sqrt{3}) \approx 0.537422$, when the spheres are placed using the underlying structure of a (non-gyrated) tetrahedral-octahedral honeycomb. That is, in this arrangement, the particles are the inscribed spheres to the octahedra comprising the honeycomb. This packing fraction is relatively close to the reported RLP value for frictional spheres in 3D, $\eta \approx 0.536$~\cite{song2008phase}. Similarly, values of $\eta$ close to where a divergence in the effective viscosity of a sphere suspension with sliding and rolling friction is reported~\cite{singh2020shear}, can be obtained by considering other congruent honeycombs. That is, a diamond lattice of triakis truncated tetraheda gives $\eta = \sqrt{3}\pi / 16 \approx 0.34009$, while Singh~\textit{et al.} report $\eta \approx 0.36$ using their simulations, see the ESI~\cite{ESI}.

Turning to the 3D hard-sphere glass transition, the literature places the transition over a wide range of $\eta$ close to $\eta = 0.6$. For example, Zaccarelli~\textit{et al.}~\cite{zaccarelli2015polydispersity} provide $\eta_{g} \approx 0.588$ using a power-law fit on the MSD the largest particles in a in a polydisperse sample --- a value that is also reported for some jamming papers~\cite{singh2020shear, dangelo2025granular} --- while an exponential fit leads to a value of $\eta_{g} \approx 0.639$. A recent work by Geiger~\textit{et al.}~\cite{geiger2024decoupling} on a bidisperse sphere mixture places the transition at $\eta_{g} \approx 0.61$. Given the spread, it seems sensible to invert the question of finding a geometric ground state based on a congruent honeycomb for a given volume fraction from the literature. A preliminary search, see the ESI~\cite{ESI}, did not reveal any candidate structure that naturally falls into this range. The full exploration of honeycomb structures as potential sources of dynamic arrest is therefore left to follow-up work. 

%%%%%%%%
\section{\label{sec:close}Closing Remarks}
%%%%%%%%

Summarizing, I have argued for the existence of geometric ground states with congruent Voronoi tiles, which induce continuous dynamic transitions in 2D systems of hard disks. The appearance of dynamic arrest then occurs in a manner that is analogous to the one found for model foams and tissues, where geometry and topology are well-known to play a key role.

This insight is based on careful analysis of the isoperimetric quotient, mean squared displacement, and percolation properties. Specifically, the system that I numerically investigated falls out of equilibrium for a nearly constant value of the area fraction $\phi \approx 0.777$ over a wide range of disk size ratios $R \lesssim 0.83$. Together with experimental evidence in both quasi-2D Brownian suspension~\cite{lozano2019active} and 2D granular systems~\cite{puckett2013equilibrating, naseer2025micromechanics, vishali2025topological}, among others, that have dynamical changes for similar values of $\phi$, this points toward a common underlying feature.

I identify this feature to be a geometric ground state, given by a floret pentagonal arrangement of disks on the plane. This configuration has an area fraction of $\phi_{g} = \sqrt{3}\pi / 7 \approx 0.777343$, which corresponds to the observed transition values. I have tested this argument for a trinary system to validate my approach and found excellent agreement, further suggesting that system composition is not important. In addition, carrying the geometric-ground-state reasoning further, it seems plausible that the experimental observations of a caging transition by Li~\textit{et al.}~\cite{li2020anatomy} can be similarly explained using a 2D honeycomb arrangement as geometric ground state. However, the evidence presented here for this correspondence is more tenuous than it is for the glass/random-loose packing transition. In 3D, the reasoning presented here can be carried over to explain the random-loose packing volume fraction for spheres using the square bipyramidal honeycomb arrangement.

Future work in this direction should focus on making the suggested connection between geometry and dynamics more mathematically rigorous. Physically, it is important to probe experimentally the constancy of the transition $\phi$ in both granular and Brownian systems, as well as elucidate which preparations lead to the predicted value of $\phi$ for the former. Thinking in terms of geometric ground states could also help shed light on attempts to explain the crossover between athermal jamming and the thermal glass transition~\cite{dinkgreve2018crossover, cao2021rheology}, although it should be clear that 3D systems can be quite different from their 2D counterparts.

%%%%%%%%
\section{\label{sec:ackn}Acknowledgements}
%%%%%%%%

The author acknowledges financial support from the Netherlands Organization for Scientific Research (NWO) through Start-Up Grant 740.018.013. I am grateful to Frank Smallenburg for making his code available for use~\cite{smallenburg2022efficient}, which proved instrumental in the realization of this project, as well as for useful discussions on the analysis. I further acknowledge Clemens Bechinger and Celia Lozano for providing data~\cite{lozano2019active}, as well as useful discussion. Lastly, I would like to thank Karen Daniels, Luca Giomi, Wilson Poon, Roel Dullens, Alptug Ulug{\"o}l, and Edan Lerner for providing useful input on my work and/or pointing out worthwhile references.

An open-data package containing the means to reproduce the figures and overall results of the simulations is available at: [\redc{url to be inserted}].


\clearpage
\appendix
\onecolumngrid

%%%%%%%%
\section{\label{sec:compare}Comparison to the Literature}
%%%%%%%%

\noindent My study into the dynamics and structure of 2D bidisperse disks builds on decades of research into the structure and dynamics of dense (quasi-)2D suspensions. In this appendix, I will compare my results to those obtained in the literature, where I aim to be as extensive as possible~\footnote{It is nearly inevitable that a literature reference has been missed; I apologize should your work unwittingly not receive the same attention as those listed in this section.}. Again, referencing the $R$-$\phi$ diagram of Fig.~\ref{fig:diagram}, I have put (estimated) data points from the literature on top of my results, labeling these ``a'' through ``r''.

\begin{figure}[!htb]
\centering
\includegraphics[width=85mm]{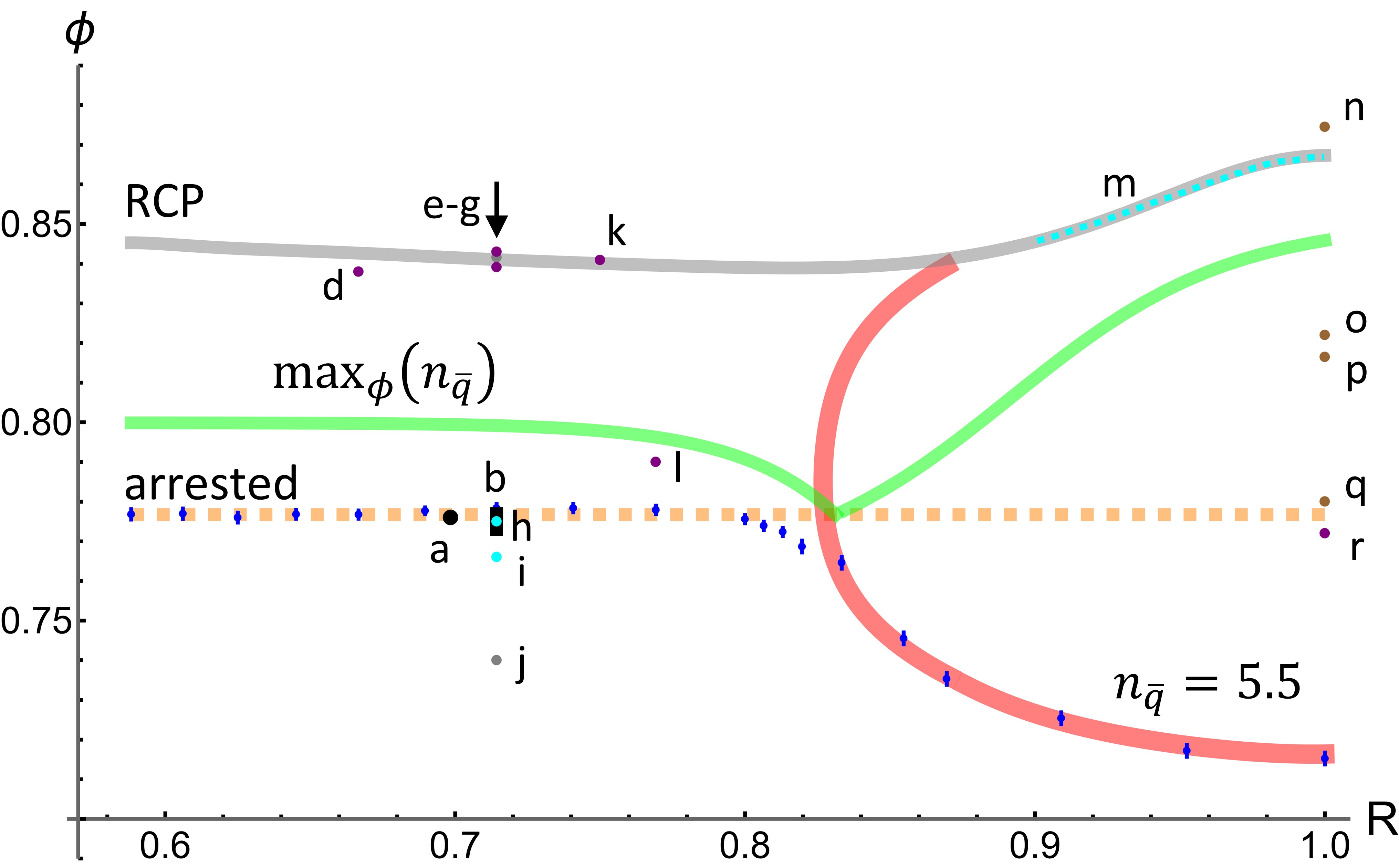}
\caption{\label{fig:compare}\textbf{Comparison to the literature.} The $R$-$\phi$ state diagram for bidisperse disks in 2D is the same as in Fig.~\ref{fig:diagram}a. The dots marked with letters (``a''-``r''; ``m'' corresponds to the dashed cyan curve) indicate systems for which literature values are known. The large black point ``a'' represents the experimental system of interest~\cite{lozano2019active} and the one marked ``b'' the one of Refs.~\cite{puckett2013equilibrating, vishali2025topological}, as before. The purple points indicate systems that can be directly mapped to my simulations. Cyan data points represent comparable systems, for which some interpretation of the mapping is required. Brown points follow from theoretical calculations and gray points represent systems that may be compared to my data, but for which the translation is not necessarily obvious. The text in this appendix provides the details on all comparisons and how to interpret these.}
\end{figure}

Point ``a'' and ``b'' are the experimental references in thermal~\cite{lozano2019active} and athermal~\cite{puckett2013equilibrating, vishali2025topological}, respectively. Point ``c'' is omitted, as I previously showed this in Fig.~\ref{fig:pdfpr} and it falls outside of the $\phi$ range shown here. There is also a point ``s'', which is covered in Appendix~\ref{sec:press}, and not shown here to keep the presentation clean. I will now go over these points and discuss how they were obtained and what can be learned from the comparison.

Points ``d''-``g'', ``i'', ``m'', and ``n'' relate to the RCP value $\phi_{m}$ obtained using the algorithm of Ref.~\cite{desmond2009random}. The first point, labelled ``d'', was obtained from the experimental study of Zhao~\textit{et al.}~\cite{zhao2012correlation}, wherein the two-point correlation of free Voronoi volumes in binary disc packings was measured. The size ratio of their teflon disks is $6$~mm to $9$~mm, which results in $R \approx 0.67$. The authors report a maximum value of $\phi = 0.838$ for their packing, which is in good agreement with the RCP value that I determined.

From top to bottom, points ``e'' through ``g'' correspond to data with $R^{-1} = 1.4$ that was obtained from Refs.~\cite{silbert2010jamming} ($\phi \approx 0.843$),~\cite{chieco2018spectrum} ($\phi \approx 0.8416$), and~\cite{atkinson2014existence} ($\phi \approx 0.8391$), respectively. The top data point derives from a numerical study into the effect of friction on the jamming properties of disordered particle packings of bidisperse disks~\cite{silbert2010jamming}. This point was obtained in the frictionless limit, which is appropriate here. The middle data point was extracted from a study of 1:1 bidisperse harmonically repulsive disks with 7:5 diameter ratio~\cite{chieco2018spectrum}. Here, I present the fitted value of the jamming fraction as extracted in Appendix~B2 of their work. Note that despite the softness of the potential this extracted jamming point agrees very well with the RCP value reported here. Lastly, the bottom data point specifically examines isostatic, maximally random jammed monodisperse hard-disk packings~\cite{atkinson2014existence}. This point too matches my $\phi_{m}$ value well, which is expected for a study into jamming.

The point labelled ``i'', derives from Ref.~\cite{rieser2016divergence}. Rieser~\textit{et al.} considered the divergence of Voronoi cell anisotropy vector for both experimental and numerical data on bidisperse disk-like packings~\cite{rieser2016divergence}. Here, I take the value $\phi \approx 0.8409$ that was obtained using their quenched simulation protocol for $R = 0.75$; the $\phi$ for their thermalized data point ($\phi \approx 0.8465$) agreed less well with the RCP value in Fig.~\ref{fig:compare}. It may be that thermalization makes the softness of the simulated particles play a more significant role.

The curve labelled ``m'', was obtained from Ref.~\cite{koeze2016mapping}, which systematically mapped out the jamming transition of all 2D bidisperse mixtures of frictionless disks in the hard particle limit. In particular, I extracted the data from their Fig.~8 for the largest number of particles simulated and interpolated this to arrive at the dashed cyan curve. Finally, point ``n'' comes from Ref.~\cite{parisi2010mean}, wherein a theory of amorphous packings and more generally glassy states of hard spheres is reviewed. Their Table~III reports a 2D value of $\phi \approx 0.8745$ that follows from a first-order small cage expansion of the `glass close-packing' density. The agreement is reasonable given the nature of the theoretical approximation employed in Ref.~\cite{parisi2010mean}.

I conclude that the RCP values obtained using algorithm of Desmond and Weeks~\cite{desmond2009random} agree well with the various pieces of (more modern) literature. Note that I also already commented on the excellent agreement between my numerical result and the experimentally obtained value of $\phi$ for which the SISF relaxation time diverges~\cite{lozano2019active} in Section~\ref{sec:dyna}. All of this together provides confidence in my choice of setting $\phi_{m}$ as an upper bound to the $\phi$ range.

Next, let us cover points ``j'', ``o'', and ``p'', which do not appear to match any feature in the state diagram, before moving onto a discussion of the remaining points that do. Point ``j'' derives from a numerical study of bidisperse systems ($R^{-1} = 1.4$) comprising soft harmonic disks in 2D~\cite{maiti2019thermal}. The authors report a ``thermal jamming transition'', wherein at $\phi \approx 0.74$ an effectively non-ergodic state sets in for which there remain overlaps. The application of this result to the hard-disk packings I have studied is slightly tenuous, however, their $\phi$ value approaches where I place the onset of apparent arrested dynamics. That being said, it much more closely matches the $n_{p} = 5.5$ percolation value discussed in Appendix~\ref{sec:perc} (not shown here). Thus, there is no obvious conclusion that can be drawn from perceived agreement and any correspondence is more than likely a fortuitous coincidence.

Point ``o'' was obtained from the combined theorical and experimental work of Ref.~\cite{bideau1986geometrical}, wherein an approximate expression for the close-packing density $\phi \approx 0.822$ of a monodisperse sample is proposed. Clearly, the point deviates considerably from the RCP value obtained here, though it agrees with a previous experimental study to within the error~\cite{berryman1983random}. I have nonetheless included this point, as the theory put forward in this source was used to obtain a data point (``q'') that agrees better with one of my measures. Lastly, point ``p'' follows from the theory of Ref.~\cite{parisi2010mean} (see their Table~III), and is identified as the point where a thermodynamic glass transition is supposed to happen. Within the limitations of my analysis, I find no evidence for this transition. However, I should stress that I did not pursue analysis of the ideal glass transition; if it is even present for a monodisperse system~\cite{donev2007configurational}.

This leaves points ``h'', ``i'', ``l'', ``q'', and ``r''. The first, point ``h'', follows from Ref.~\cite{donev2007configurational}, wherein a binary hard-disk mixture is studied with ratio of disk diameters $R^{-1} = 1.4$. The authors report $\phi \approx 0.775$ for the (equilibrium) freezing area fraction. However, this study was performed for a 1:2 stoichiometric ratio of large to small disks. I will argue in Section~\ref{sec:geound} how this fits my interpretation of $\phi_{a}$.

Next, point ``i'' follows from Ref.~\cite{silbert2010jamming}, which studied frictional hard-disks with $R^{-1} = 1.4$. Here, I fitted the jamming data points with a sigmoidal curve and extrapolated to the infinite friction result to a RLP value of $\phi \approx 0.766$. The reason for taking this fitting approach is that Silbert does not report the 2D jamming value. Moreover, there is a peculiar nonmonotonicity to the coordination-number data for 2D systems provided in Table~I of Ref.~\cite{silbert2010jamming}. Given the level of uncertainty on the data provided, the agreement between my fitted RLP value of $\phi \approx 0.766$ and the arrested-dynamics area fraction $\phi_{a}$ obtained in this study is very good. It is curious that there appears to be a connection between RLP of a fully frictional sample and the apparent arrested dynamics in a frictionless system. My interpretation is that this is another signal of an underlying geometric feature, see Section~\ref{sec:geound}.

Point ``l'' was obtained from Ref.~\cite{zhang2024anisotropic}, the authors of which took a very similar approach to studying 2D systems with arrested dynamics. They considered the evolution of Voronoi cell geometry in simple hard-disk models by simulations and colloid experiments. In particular, they located the glass transition using a Voronoi-based (neighborhood) cage anisotropy parameter. Their anisotropy takes the ratio of the long and short axes of the Voronoi cells and has a peak similar to what I find when studying $\max_{\phi} \! \left( n_{\bar{q}} \right)$. Here, I use their point $\phi \approx 0.79$ which was obtained \textit{via} 2D EDMD of bidisperse disks with $R^{-1} = 1.3$ and a 1:1 number ratio. This data point is in good agreement with where I locate $\phi_{a}$ and $\max_{\phi} \! \left( n_{\bar{q}} \right)$, respectively, given the subtle differences in approach and the uncertainty this imposes. However, based on my analysis, a change in cage anisotropy does not fully coincide with the apparent kinetic arrest, though it is possible that a measure based on aspect ratio, as pursued in Ref.~\cite{zhang2024anisotropic}, leads to a closer connection.

Reference~\cite{hinrichsen1990random} makes use of the theory by Bideau~\textit{et al.}~\cite{bideau1986geometrical} to predict the RLP density of $\phi \approx 0.78$, which is indicated by point ``q'', for a monodisperse sample of hard disks. The point ``r'' follows from the numerical study performed in the same paper, which places the RLP area fraction at $\phi \approx 0.772$. Given the limitations of this study, the agreement with the dashed green line, and thus the minimum in the amount of pentagonal large-particle neighborhoods, is excellent. This is a second piece of evidence --- also see point ``i'' above --- in the direction of geometric features underlying the $\phi \approx 0.777$ line in the systems considered in my research. Note that Ref.~\cite{hinrichsen1990random} reports on an average number of contacts per disk being $\approx 3$ at their RLP $\phi$. Using my Voronoi approach, I find an average number of neighbors that is substantially larger, see Fig.~\ref{fig:pdfs}d. It is, however, interesting to assess how many of these neighbors contribute to `immobilizing' contacts, such as considered in Ref.~\cite{hinrichsen1990random}, at any one time.

Lastly, a recent work on the ideals glass transition in 2D disks for a polydisperse systems reports $\phi \approx 0.78$~\cite{bolton2024ideal}, which is again close to the FPT value that I obtained. The polydisperse nature of the system provides further (circumstantial) evidence for an underlying geometric feature to the transition.

%%%%%%%%
\section{\label{sec:frac}Pseudo-Polygon Fractions}
%%%%%%%%

\begin{figure*}[!htb]
\centering
\includegraphics[width=175mm]{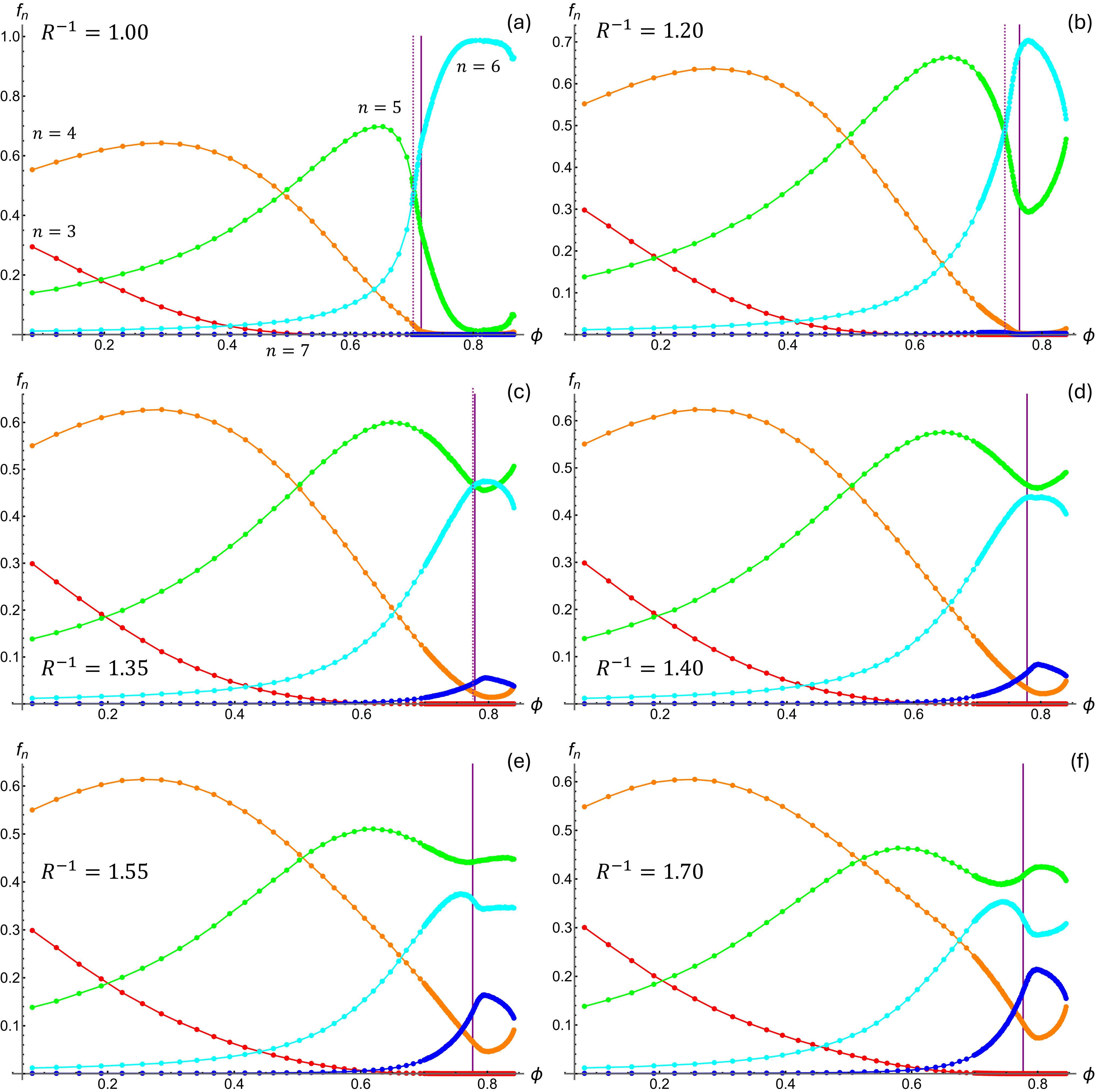}
\caption{\label{fig:fracts}\textbf{Fractions of pseudo-polygons in the sample.} (a-f) The fraction $f_{n}$ of pseudo-triangles (red, $n=3$), -squares (orange, $n=4$), -pentagons (green, $n=5$), -hexagons (cyan, $n=6$), and -heptagons (blue, $n=7$), as a function of the area fraction $\phi$. The dots show the sampling points, and the curves are guides to the eye that connect these. The relevant inverse size ration $R^{-1}$ is indicated in each panel. The vertical solid purple line indicates the value of $\phi$ for which the system falls out of equilibrium, $\phi_{a}$,~\textit{i.e.}, where the extrapolated diffusion coefficient vanishes. The vertical dotted purple line indicates the value of $\phi$ for which there is a larger fraction of pseudo-hexagons than there is of pseudo-pentagons.}
\end{figure*}

\noindent In the main text, I analyze features of $P_{\phi}(q)$, see Fig.~\ref{fig:pdfs}, to establish features of the state diagram, see Fig.~\ref{fig:diagram}. One aspect of this analysis is the crossover between the highest mode (of the PDF) being pentagonal to hexagonal, see Fig.~\ref{fig:pdfpr}. This may give the false impression that the system transitions from being mostly pentagonal to mostly hexagonal,~\textit{i.e.}, undergoes a disordered-to-ordered transition. Here, I therefore examine the fraction of pseudo-polygons, see Fig.~\ref{fig:fracts}, which shows the fraction of pseudo-$n$-gons $f_{n}$ as a function of the area fraction $\phi$.

My study of the fraction of pseudo-polygons reveals that for low $\phi$, there is an abundance of low-$n$ neighborhoods,~\textit{i.e.}, pseudo-triangles and -squares. This makes sense, as the neighborhoods are relatively independent of the particle shape. As $\phi$ increases, the fraction of pseudo-pentagons and pseudo-hexagons rises. The first takes over in highest abundance at $\phi \approx 0.5$, while the latter can become the dominant shape for sufficiently high values of $R$ only. I further note that as $\phi$ increases to its RCP value $\phi_{m}$, any trend of increase in $f_{n}$ transitions toward one of decrease and \textit{vice versa}. This can be explained by the fact that in order to compact the system maximally in the square simulation boxes that I used throughout the study, defects must be introduced for the hexagonal crystal structures. The situation is less clear cut for the systems that (appear) to remain amorphous. This lowers the abundance of the prevalent pseudo-polygon and can increase fractions of less abundant neighborhood shapes.

For $R^{-1} > 1.35$ (equivalently $R \lesssim 0.74$), I find that for all $\phi$ the inequality $f_{6} < f_{5}$ holds (vertical dotted purple lines in Fig.~\ref{fig:fracts}). Note that $R \approx 0.74$ is not value for which the $n_{\bar{q}} = 5.5$ curve closes in on itself, which occurs for $R \approx 0.83$. Thus, there is no transition from pentagonal to hexagonal below $R \approx 0.74$, which presumably changes the behavior of the system from being polycrystalline to being truly disordered. However, as is evidenced by the neighborhood composition graphs in Fig.~\ref{fig:fracts}, the system takes on a range of compositions as a function of $\phi$ even for relatively low $R$. So one has to be careful in drawing to strong conclusions based on Fig.~\ref{fig:fracts}.

Lastly, Fig.~\ref{fig:fracts} visualizes where there is kinetic arrest using a vertical solid purple line. In panels (a) and (b), there appears to be a subtle kink in the $n=6$ and $n=5$ data, respectively, where I locate $\phi_{a}$. However, also considering the other data, I was unable to find an obvious feature to the fraction of pseudo-$n$-gons data that correlates with the apparent kinetic arrest. If it is present in this data, it is merely a subtle effect.

%%%%%%%%
\section{\label{sec:press}Crystallinity and Pressure}
%%%%%%%%

\begin{figure*}[!htb]
\centering
\includegraphics[width=175mm]{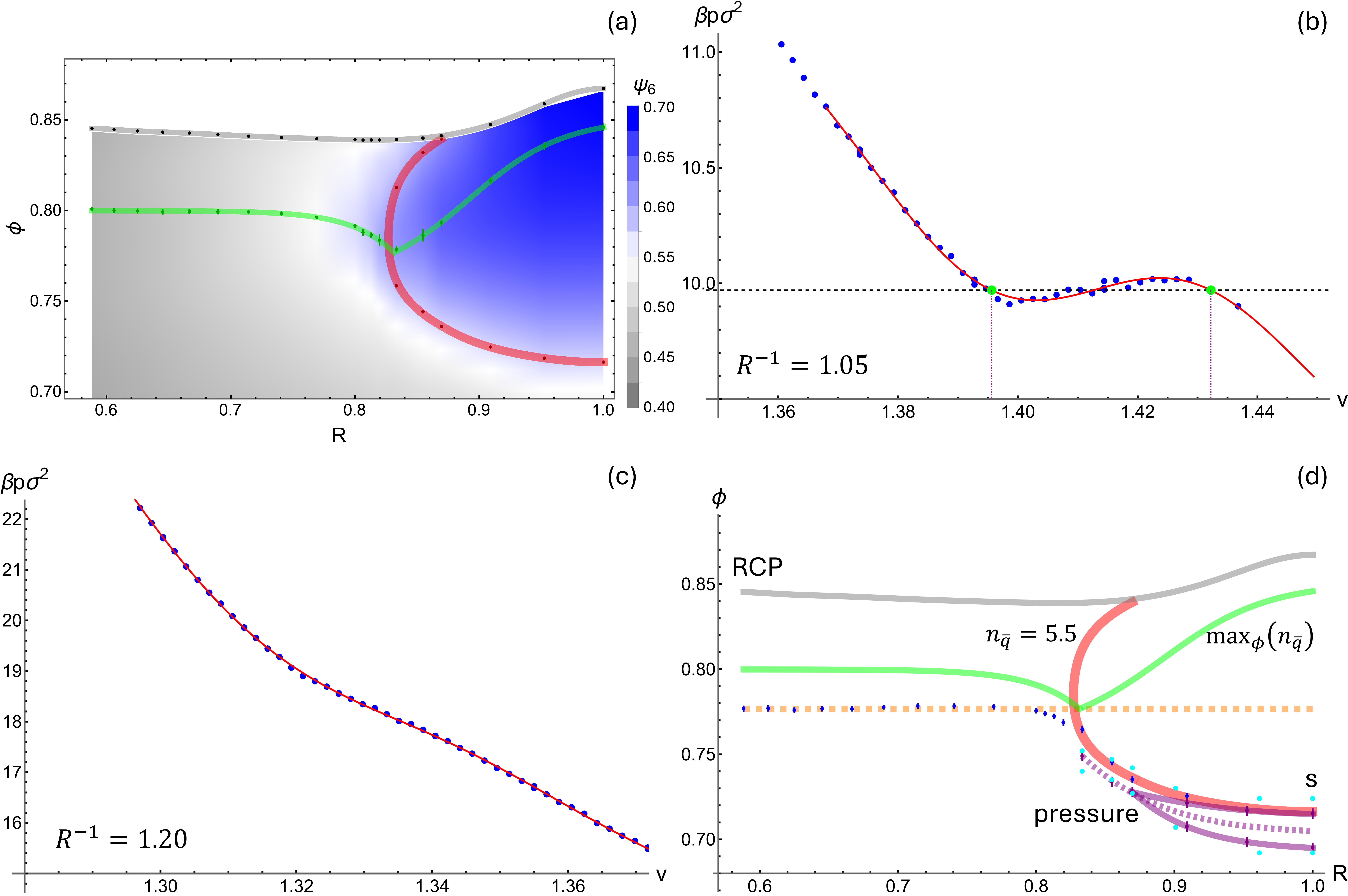}
\caption{\label{fig:press}\textbf{Determining the phase transition via the Steinhardt order parameter and by measuring the pressure.} (a) The order parameter $\psi_{6}$ indicated in the state diagram of Fig.~\ref{fig:diagram}, the use of color is indicated by the legend. Only the RCP (black), the $n_{\bar{q}} = 5.5$ (red), and the  $\max_{\phi} \! \left( n_{\bar{q}} \right)$ (green) curves are shown from the data presented in Fig.~\ref{fig:diagram}. (b-c) Pressure curves (blue points) showing the reduced pressure $\beta p \sigma^{2}$ --- $\beta$ is the inverse thermal energy and $\sigma$ is the diameter of the largest particle --- as a function of the inverse particle density $v = V/N = \rho^{-1}$. The relevant size ratio $R$ is provided in the bottom-left corner of the graphs. The red curves represent polynomial fits to the data. (a)~This graph shows a clear van-der-Waals loop for which the coexistence pressure is indicated by the horizontal dashed black line. The two extracted coexistence points are indicated in green and the associated $v$ using vertical dotted purple lines. (b)~This graph shows an inflection point, but no longer a van-der-Waals loop. (c)~The $R^{-1} = 1.2$ data still has a change in slope, for which the minimum slope can be fitted. (d)~The $R$-$\phi$ state diagram of Fig.~\ref{fig:diagram} with the $\phi$ for which there is dynamic arrest indicated in blue. On top of this, I have added the pressure data extracted from the van-der-Waals loops and changes in slope in purple. The solid purple curves guide the eye for the coexistence area fractions, while the dashed purple curve guides the eye for the minimum-slope data below $R \approx 0.87$. Below $R \approx 0.82$ I was unable to accurately determine any such minimum. Error bars denote the standard error of the mean in all cases. The cyan data points labelled ``s'' are obtained from Ref.~\cite{huerta2012towards}, wherein coexistence $\phi$ and $p$ were determined.}
\end{figure*}

\noindent In this section, the pressure and bond-orientational order parameter are considered. Figure~\ref{fig:press}a compares the $n_{\bar{q}} = 5.5$ curve and the behavior of the traditional Steinhardt order parameter $\psi_{6}$. Smoothing and interpolation had to be applied to achieve the surface plot, see the ESI~\cite{ESI} for details. The literature provides reasonable average values for $\psi_{6}$~\cite{tamborini2015correlation}, which the data in Fig.~\ref{fig:press}a adheres to. However, to the best of my knowledge, clear dividing values for fluid and (locally) crystalline configurations, as are available for 3D, are lacking. Here, I assume that $\psi_{6} \approx 0.54$ is a good divide for when crystalization occurs, based on the $\phi$ values for which a monodisperse system is known to crystallize. This gives agreeable correspondence with the $n_{\bar{q}} = 5.5$ curve.

The is coexistence between a disordered fluid, hexatic phase, and hexagonal crystal at values of $R \approx 1$ has attracted considerable historic interest~\cite{strandburg1988two, gasser2009crystallization, engel2013hard, qi2014two, kapfer2015two}. Given the system size that I considered, picking up on the hexatic phase is not realistic~\cite{engel2013hard, qi2014two}. However, there should be features of a first-order transition between a disordered fluid and ordered hexagonal phase in my data.

In the canonical ensemble, there is a van-der-Waals (vdW) loop in the pressure $p$ as a function of the inverse particle density $v = V/N = \rho^{-1}$. I therefore computed the pressure from the collisions in the system \textit{via} the momentum exchange during collisions, using the standard virial expression. This was averaged and subsequently used to establish vdW loops whenever they are present, see Fig.~\ref{fig:press}b for an example of this structure. For sufficiently low values of $R$,~\textit{i.e.}, values of the size ratio below $R \approx 0.87$, I was unable to detect a loop structure. However, my data was still of sufficient quality to establish a minimum slope. That is, as the loop vanishes, there is an inflection point, suggestive of a critical point, see Fig.~\ref{fig:press}b.

Lowering $R$ further (up to $R \approx 0.82$), I still observe a change in slope, see Fig.~\ref{fig:press}c, for which I can determine the value of $v$ (or equivalently $\phi$) where the slope attains a local minimum. Figure~\ref{fig:press}d shows the coexistence points and local minima obtained using the pressure curves on top of the state diagram from Fig.~\ref{fig:diagram}. I observe that for $R$ close enough to $1$ the top coexistence curve matches the pseudo-polygon curve $n_{\bar{q}} = 5.5$ quite well. The agreement is less convincing for lower $R$, but this is an additional piece of information rather than the main goal of the study. Given the intrinsic difference in ensemble, no attempt was made to improve the quality of the data.

Lastly, I compared my coexistence data to that reported in the literature~\cite{huerta2012towards}, which was obtained using a standard Monte Carlo simulation technique based on the Metropolis algorithm. The number of particles considered in this study is smaller (400), yet I observe an acceptable agreement between their results and mine for $R \approx 1$. For lower $R$, I do not pick upon coexistence, presumably because there is a difference in the way the studies are set up.

%%%%%%%%
\section{\label{sec:perc}Shell and Shape Percolation}
%%%%%%%%

\noindent In order to confirm the nearly constant value of $\phi_{a}$ in the range $R \lesssim 0.8$, also see Fig.~\ref{fig:diagram}, I considered percolation in the system. When working with hard particles, contact percolation only occurs at RCP. However, when there is a substantial structural change, or the system falls out of equilibrium, there should be a trace of this in the number of collisions. As the temperature --- the average particle velocity --- is maintained constant, physical intuition suggests that the gaps between particles should reflect such a change in collision rate. This was analyzed as follows.

\begin{figure*}[!htb]
\centering
\includegraphics[width=175mm]{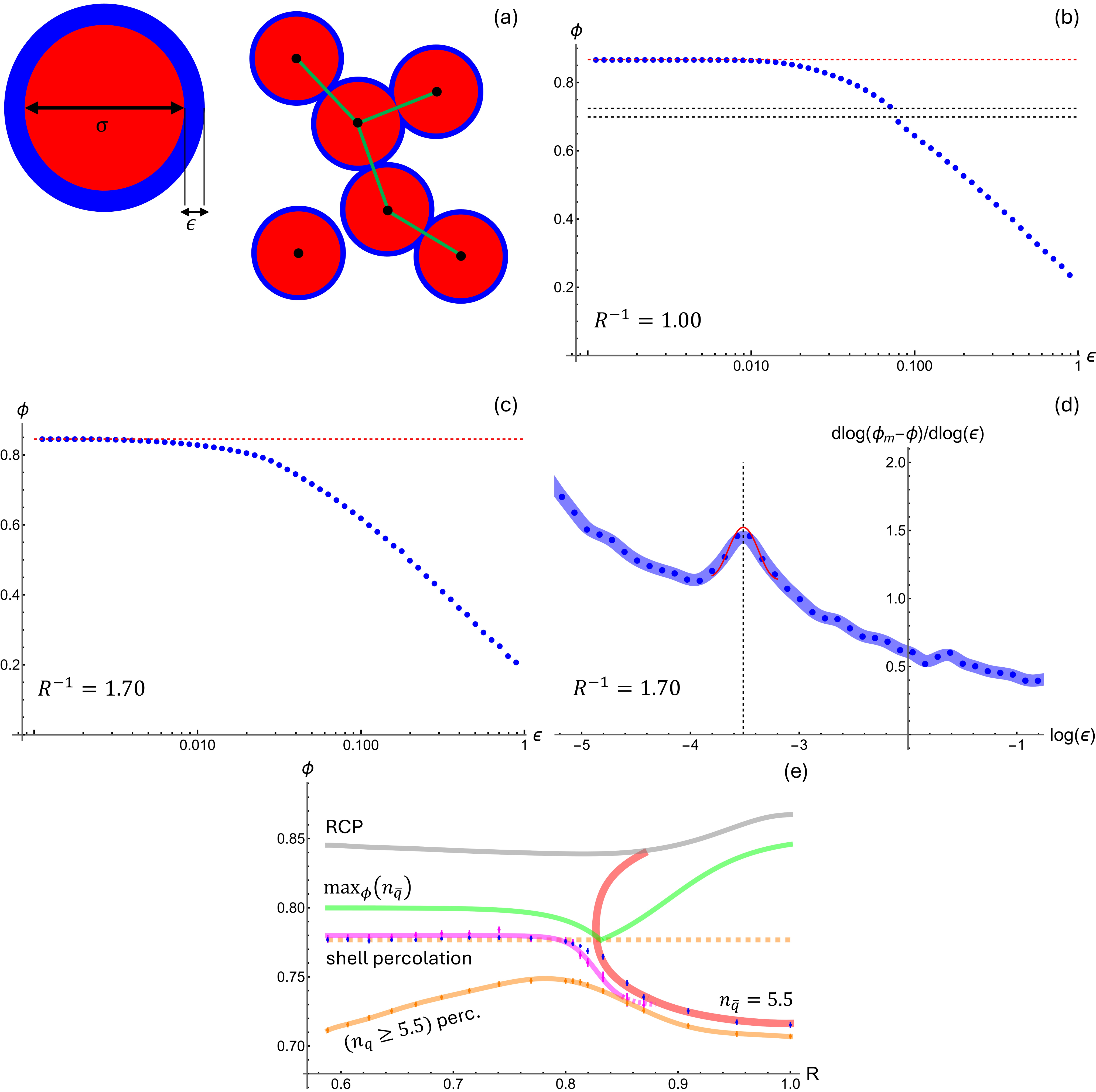}
\caption{\label{fig:isoper}\textbf{Percolation transitions based on contact and isoperimetric quotient.} (a)~Illustration of the shell-based percolation procedure. Each hard-disk particle (red) with diameter $\sigma$ is surrounded by a shell (blue) of width $\epsilon$, which does not impact the structure. These virtual shells are used to determine local clusters, as illustrated for a small cluster of four particles using green lines. Only a monodisperse example is shown, but the procedure is equivalent for a bidisperse system. (b,c)~The value of the area fraction $\phi$ for which the percolation transition occurs, as a function of the shell width $\epsilon$ in log-linear representation. The dashed red line indicates the asymptotic value for $\phi$ when $\epsilon \downarrow 0$, which corresponds to the RCP value $\phi_{m}$ shown in Fig.~\ref{fig:diagram}. In panel (b), the system is monodisperse ($R = 1$) and there appear to be two branches to the curve, as indicated using the dotted black lines, which guide the eye to $\phi = 0.699$ and $\phi = 0.724$; the region over which there are transitions~\cite{engel2013hard, qi2014two}. In panel (c), the system is bidisperse with an inverse size ratio $R^{-1} = 1.7$. Note that the curve appears relatively smooth. (d)~Approach to finding the value of $\epsilon$ for which the system changes. The curve shows the numerically computed value of $( \partial \log \left[ \phi_{m} - \phi \right] )/( \partial \log \epsilon )$, which shows a clear peak near $\log \epsilon = - 3.5$. This peak is fitted using a fourth-order local polynomial fit (red curve) and used to establish the $\epsilon$ value for which the system undergoes a change, $\epsilon^{\ast}$, as indicated using the dotted black line. This corresponds to an area fraction of $\phi \approx 0.777$. (e)~Representation of the $R$-$\phi$ diagram in Fig.~\ref{fig:diagram}, which shows the $\phi^{\ast}$ following from the shell percolation $\epsilon^{\ast}$ in magenta. The blue data shows the kinetic arrest obtained from the MSDs. Another form of percolation was obtained by establishing when a system-spanning cluster of local neighborhoods with $n_{q} \ge 5$ was obtained (orange data and solid orange curve). That is, a cluster of particles whose neighborhood is closer to being hexagonal than it is to being pentagonal. The dots indicate the data, the error bars show the standard error of the mean, and the curves guide the eye.}
\end{figure*}

The size of each particle was augmented with a small shell of size $\epsilon$, that is, when particle $i$ has a diameter $\sigma_{i}$, it is treated as having a diameter $\sigma_{i} + 2\epsilon$, see Fig.~\ref{fig:isoper}a. Next, I used a cluster algorithm to identify the largest cluster and establish whether the system is percolating by checking whether it spans the simulation box in both directions~\cite{bug1985interactions, lebovka2024percolation}, see the ESI~\cite{ESI} for the code. This is done over all realizations of the system at a given $R$ and $\phi$.

Using this data, the fraction of times that the system is clustered can be established. Between $\phi = 0$ and the RCP value $\phi_{m}$, I always found some fraction of the systems to clusters and this $\phi$-dependence can be fitted accurately using a sigmoid~\footnote{In the limit of an infinite system size, the sigmoid becomes a Heaviside function about the percolation $\phi$.}, see the ESI~\cite{ESI}. The mid-point of the sigmoid is used to establish the percolation area fraction $\phi(\epsilon)$, as shown in Fig.~\ref{fig:isoper}b,c; these are two representative samples of the full data.

\FloatBarrier

It is clear that the percolation area fraction tends asymptotically toward the RCP value $\phi_{m}$ for $\epsilon \downarrow 0$. Interestingly, there are two branches to this shell-percolation $\phi$, when the system undergoes a (first-order) phase transition from a disordered fluid to a hexagonal crystal, see Fig.~\ref{fig:isoper}b for $R = 1$. While the two branches are visually obvious, it proved difficult to fit the end points of the two branches with sufficient accuracy.  The correspondence with known phase transition results is therefore only indicated visually in Fig.~\ref{fig:isoper}b. The data appears smooth when the system does not seem to undergo a first-order phase transition, see Fig.~\ref{fig:isoper}c for $R^{-1} = 1.7$. For such data sets, it proved possible to extract the $\phi_{a}$ by measuring the distance to $\phi_{m}$ and take the natural logarithm ($\log$) of this separation $(\phi_{m} - \phi)$ as well as that of $\epsilon$. From this data I could numerically determine the local power-law exponent $( \partial \log \left[ \phi_{m} - \phi \right] )/( \partial \log \epsilon )$ using central differences. This local power-law exponent exhibits a clear peak, see Fig.~\ref{fig:isoper}d, which was fitted using a fourth-order polynomial $a + b \left( \log \epsilon \right)^{2} + c \left( \log \epsilon \right)^{4}$ (red curve).

The peak position $\epsilon^{\ast}$ in turn was used to establish the equivalent area fraction $\phi(\epsilon^{\ast})$. Note that this position corresponds to a change in slope in the original $\phi(\epsilon)$ data. The corresponding area fraction is shown in Fig.~\ref{fig:isoper}e for those points where the was no clear indication that there were two branches (magenta data). It is directly apparent that the data accurately fits $\phi_{a}$ (blue data) for $R \lesssim 0.80$. For the remaining range, $R \gtrsim 0.81$, the trend is closer to the one of the $n_{\bar{q}} = 5.5$ curve (red).

Lastly, I studied whether there was any significance to shape percolation. That is, I note that crystallization is accurately predicted using $n_{\bar{q}} = 5.5$, but that this is a poor predictor of $\phi_{a}$. Therefore, I sought to establish those $\phi$ for which percolation is obtained when particles are added to a cluster whenever $n_{q} \ge 5.5$. This data is shown using orange in Fig.~\ref{fig:isoper}e. The last bit of the guide to the eye is dashed because it was not clear whether there was a branched or a continuous $\phi(\epsilon)$ curve. For $R \gtrsim 0.83$, the curve lies half-way between the two coexistence values of $\phi$ as established using the pressure calculation in Fig.~\ref{fig:press}. These are not shown here to avoid cluttering the figure. I conclude that there is limited value to this measure.

%%%%%%%%%%%
\section{\label{sec:median}Median Isoperimetric Quotient}
%%%%%%%%%%%

\noindent I have shown that the system crystallizes at $n_{\bar{q}} = 5.5$. However, a connection between \textit{global} properties of the neighborhood distribution and $\phi_{a}$ was less readily apparent.

\begin{figure}[!htb]
\centering
\includegraphics[width=85mm]{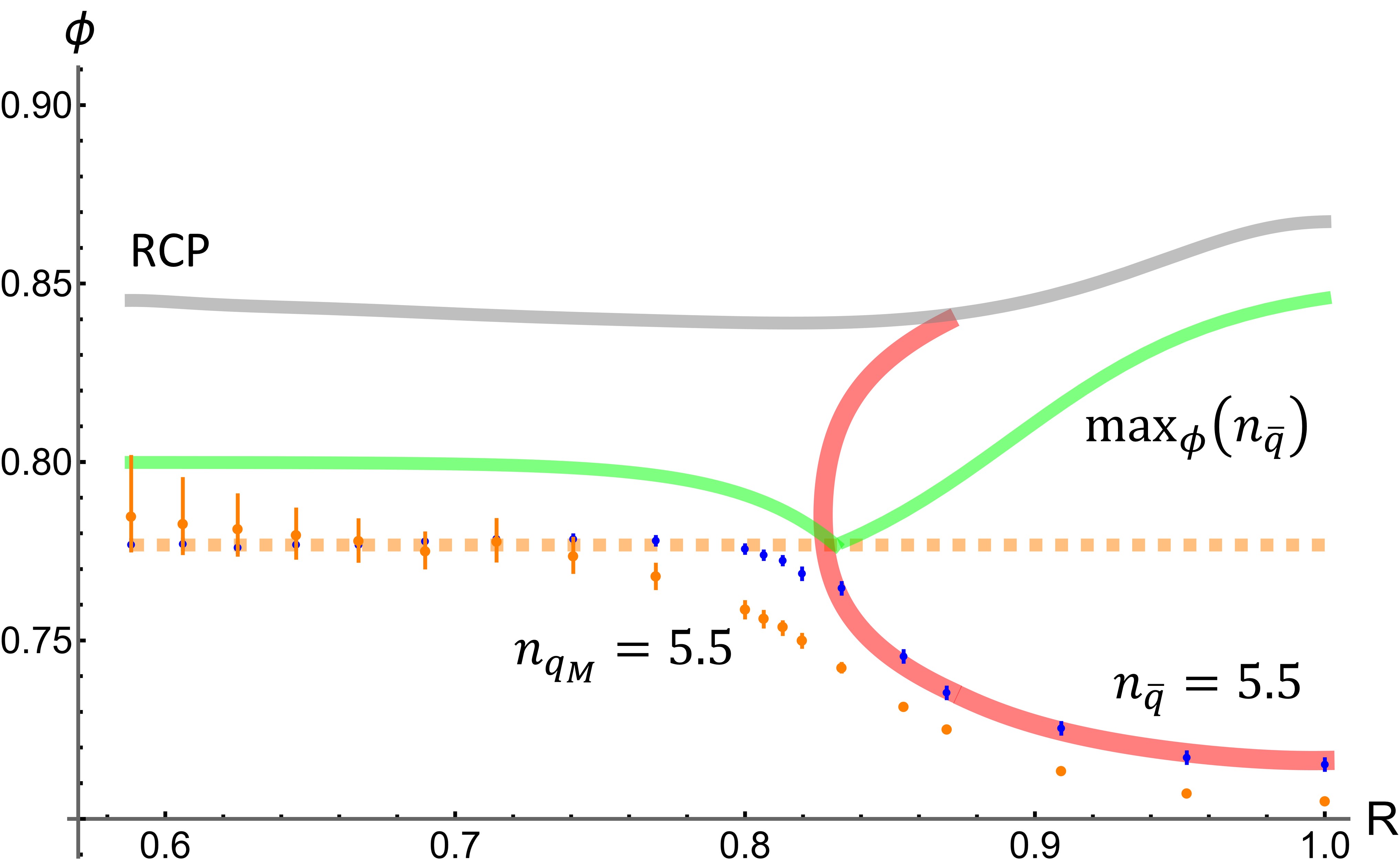}
\caption{\label{fig:median}\textbf{Properties of the median of the isoperimetric quotient distribution.} The values of $\phi$ for which the median to $P_{\phi}(q)$, $q_{M}$, has $n_{q_{M}} = 5.5$ (orange). This data is presented on top of a part of the $R$-$\phi$ state diagram of Fig.~\ref{fig:diagram}. The error bars indicate the standard error of the mean, which here is asymmetric.}
\end{figure}

Here, I therefore consider the median of $P_{\phi}(q)$, $q_{M}$, which is the $q$ value that splits PDF into two parts of equal area. This choice is further motivated by the various fraction of pseudo-polygons reported in Appendix~\ref{sec:frac}. In Fig.~\ref{fig:median}, I show those values of $\phi$, for which $n_{q_{M}} = 5.5$. Because of the error on the isoperimetric-quotient distribution, there is a sizeable level of uncertainty when it comes to the placement of the median-derived $\phi$ value, see the ESI~\cite{ESI} for additional details. However, note that for $R \lesssim 0.72$ there is reasonable agreement between the ($n_{q_{M}} = 5.5$) area fraction and $\phi_{a}$ for arrested dynamics, given the uncertainty in the data.

%%%%%%%%%%%
\section{\label{sec:fluctuations}Fluctuations of the Particle Neighborhood}
%%%%%%%%%%%

\begin{figure}[!htb]
\centering
\includegraphics[width=85mm]{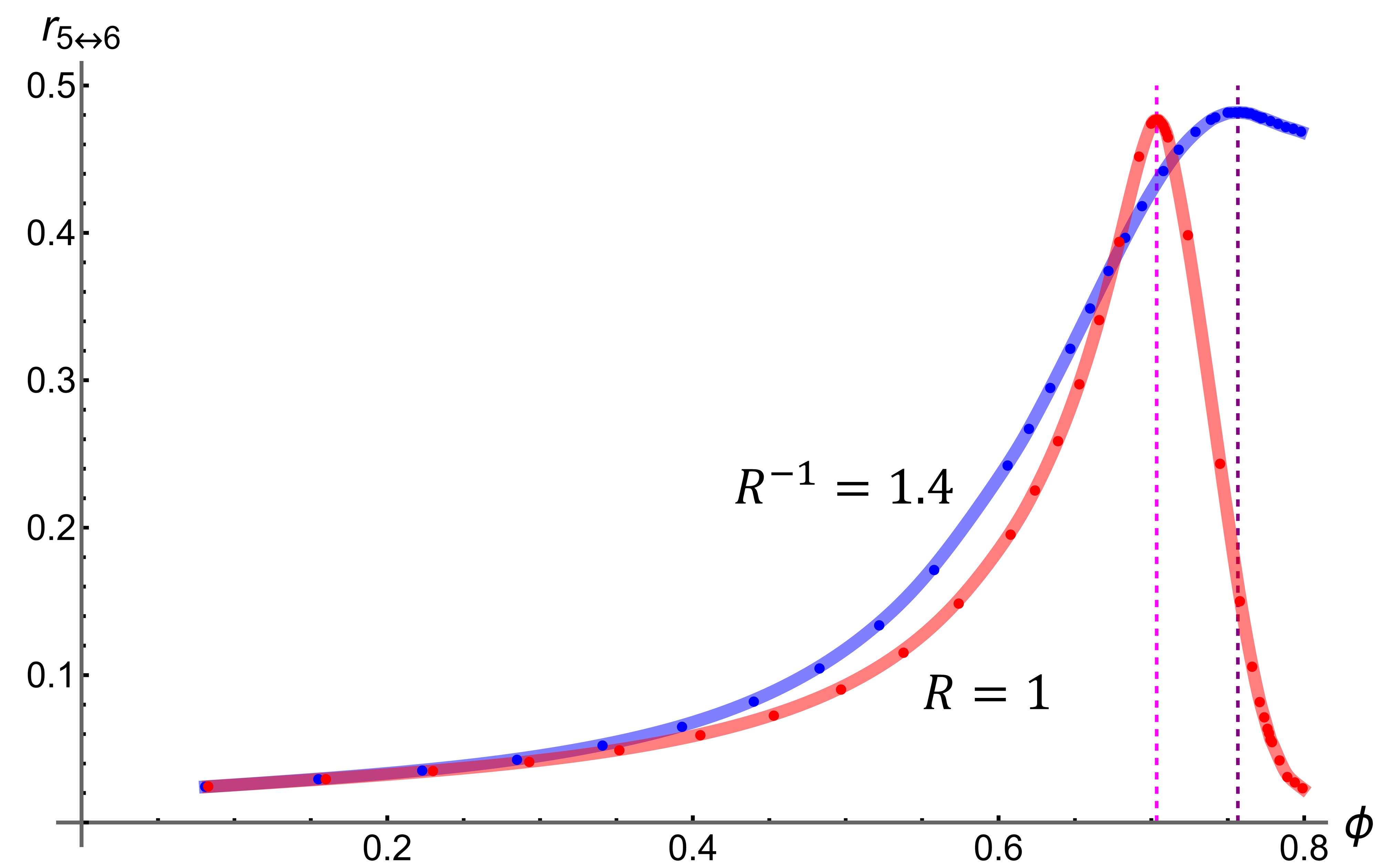}
\caption{\label{fig:trans}\textbf{The relative neighborhood exchange rate indicates strong fluctuations close to transitions.} The ratio of the number of changes between a pentagonal and a hexagonal neighborhood to the total number of particle displacements in the system, $r_{5 \leftrightarrow 6}$, as a function of the area fraction $\phi$. The red and blue points show data gathered for $R = 1$ and $R^{-1} = 1.4$, respectively, as labelled. The solid, light-red and light-blue curves serve to guide the eye. The peak positions are indicated by the magenta and purple vertical dashed lines, respectively. Errors are smaller than the symbol size and therefore not shown.}
\end{figure}

\noindent In this appendix, the frequency by which particle neighborhoods changes into one another is analyzed. The goal of this is to connect to the observations of Narinder~\textit{et al.}~\cite{narinder2022understanding}, where structure and fluctuations of a length-wise polydisperse fluid of colloidal rods was reported. In brief, for each time step, I computed the radical Voronoi diagram and kept track of the $q$ value belonging to each particle. I found that for sufficiently high $\phi$, the neighborhoods dwell about integer values of $n_{q}$ and the system does not favor half-integer $n_{q}$. For example, the trajectory of a certain particle may lead it to explore pentagonal ($n_{q} \approx 5$) and hexagonal ($n_{q} \approx 6$) neighborhoods, but the particle spends little time in $5.5$-gonal neighborhoods~\footnote{This observation does \textit{not} conflict with the fact that in Fig.~\ref{fig:pdfs}a-c, I observe dips in the $P_{\phi}(q)$ about integer values of $n_{q}$ and a smooth increasing trend near half-integer values of $n_{q}$. There is a bias to the representation of the system using $P_{\phi}(q)$, as explained in Section~\ref{sec:struct}.}. Next, I extracted transition rates for the exchange between neighborhoods and normalized this by the total number of particle displacements in the system. This resulted in, for example, relative exchange rates between pentagonal and hexagonal neighborhoods, $r_{5 \leftrightarrow 6}$, some which are shown in Fig.~\ref{fig:trans}.

Surprisingly, the relative number of exchanges constitutes a substantial fraction of particle displacements even at high $\phi$ close to $\phi_{a}$ or equivalently $\phi_{g}$. This means that while there is relatively limited cage hopping, as evidenced by the low value of $D(\phi)$, the cages themselves are quite dynamic. Additionally, I find no significant difference in the $r_{i \leftrightarrow i+1}$ between small and large particles for any integer $i$, see the ESI~\cite{ESI}. This implies that the cage dynamics is shared equally among the large and small particles, though it does not exclude a difference in cage-hopping rate,~\textit{i.e.}, in the particles' respective diffusivity.

Both data sets presented in Fig.~\ref{fig:trans} show a peak. When there is crystallization, as is the case for $R = 1$, it is sharp and very pronounced. Fitting the peak values results in $\phi \approx 0.703$ and $\phi \approx 0.757$ for $R = 1$ and $R^{-1} = 1.4$, respectively. The former agrees very well with where the fluid state is expected to become unstable. The latter is close to, but not quite at $\phi_{g}$. For $R^{-1} = 1.4$, the peak is also broader and less pronounced. Within the error, I detected no discernable drift toward one type of neighborhood as a function of time.

That the rate of cage rearrangement is relatively high, fits with the conceptual picture of a self-propelled probe particle (experiments in Ref.~\cite{lozano2019active, narinder2022understanding}) picking up on the fluctuations of the local environment, as put forward in recent modeling work from my group~\cite{abaurrea2020autonomously, narinder2022understanding}. The fact that $r_{5 \leftrightarrow 6}$ is asymmetric about $\phi \approx \phi_{g}$ would also fit with the asymmetric probe response observed in Ref.~\cite{lozano2019active}. However, there is a slight mismatch between the peak value of $\phi$ and $\phi_{g}$. This should be addressed in future work, wherein self-propelled probes are included in the sample, but this falls outside of the scope of the present research.

%%%%%%%%%%%
\section{\label{sec:3Dper}Perspective on 3D Hard-Sphere Systems}
%%%%%%%%%%%

\noindent Lastly, I will attempt to apply the notion of a geometric ground state to 3D hard-sphere systems, for which the volume fraction $\eta$ is the control parameter. To the best of my knowledge, no experiments similar to the ones performed by Lozano~\textit{et al.}~\cite{lozano2019active} and Li~\textit{et al.}~\cite{li2020anatomy} have been performed in 3D. Given that Lozano~\textit{et al.}~placed the glass transition at $\phi \approx \phi_{g} \approx 0.776$, it makes sense to try to connect to observations on the 3D glass transition, as well as 3D jamming in view of the 2D experiments of Refs.~\cite{puckett2013equilibrating, naseer2025micromechanics, vishali2025topological}. Acknowledging that the dynamics in 3D may be very different from 2D, this at least provides a reference point.

For jamming of 3D granular matter, the transition value is dependent on the details of the interaction. For frictionless, monodisperse hard spheres, the value for the RCP (frictionless jamming) is typically reported to be $\eta_{j} \approx 0.64$~\cite{rintoul1996computer, torquato2000random, mewis2012colloidal}, while for frictional spheres, the jamming point lies at $\eta_{j} \approx 0.555$~\cite{onoda1990random, silbert2010jamming, guy2015towards, anzivino2023estimating}. Singh~\textit{et al.}~\cite{singh2020shear} studied 3D jamming for no ($\eta_{j} \approx 0.648$), sliding ($\eta_{j} \approx 0.570$), and sliding \& rolling ($\eta_{j} \approx 0.365$) friction using numerical simulations and obtained three branches. It is clear from these numbers that making the connection between dynamical arrest in a Brownian suspension and RLP in a granular system is considerably more tentative.

Nonetheless, assuming that it is possible to focus on monodisperse spheres, the following analogy to the 2D hard-disk findings reported in the main text can be drawn. These should be arranged in such a way that their Voronoi cells tile space, are congruent, and result from removing particles from the closest crystalline packing \textit{via} T2 transitions. The closest packing is the face-centered cubic (FCC) arrangement, which has Voronoi cells shaped like a rhombic dodecahedron and $\eta_{c} = \pi/\sqrt{18} \approx 0.740480$. However, spheres may also be arranged in a hexagonal close packing (HCP), which has an identical packing fraction $\eta = \eta_{c}$, but the associated Voronoi cell is a trapezo-rhombic dodecahedron has two types of faces. It is well known that the FCC lattice is thermodynamically preferred over the HCP lattice, although the difference in free energy per particle between the two is small~\cite{frenkel1984new}.

Removing the central particle out of a neighborhood in an FCC arrangement, in the same spirit as done for the creation of the FPT in 2D, gives a volume fraction $\eta = 12 \eta_{c} / 13 = 2 \sqrt{2} \pi / 13 \approx 0.683520$ and associated Voronoi cells that are rhombic dodecahedron with one face extended into a point. Removing a central particle in an HCP arrangement will give two types of Voronoi cell, as the base element for HCP is a trapezo-rhombic dodecahedron, which has both trapezoids and rhombi as its sides. Both of these cells have a local volume fraction identical to that of the FCC arrangement and do not provide additional possibilities for matching. Potentially, this lack of uniform congruency upon T2 removal speaks in favor of the stability of the FCC arrangement over the HCP arrangement, but this is speculative. Turning to the body-centered cubic (BCC) lattice, I also gain two types of Voronoi cell by removing the `body' from the body-centered cubic arrangement. There are 6 next-nearest neighbors for which the square face is extended, leading to a Voronoi cell with $\eta = 6 \sqrt{3}\pi / 49 \approx 0.666294$. Additionally, there are 4 nearest neighbors for which the hexagonal face is extended, and I find $\eta = 4 \sqrt{3} \pi / 35 \approx 0.621874$. Taking the mean of the two leads to a value $\eta \approx 0.644$, which is close to the random-close packing value for hard spheres in 3D~\cite{berryman1983random}. Though this seems hard to justify, also given that generally RCP values seem to strongly depend on $\phi$ (in 2D) and $\eta$ (in 3D).

Relaxing the requirement for removing particles, gives another congruent 3D tiling. This is generated by the diamond lattice and its Voronoi cells are triakis truncated tetrahedra. This lattice is associated with a volume fraction $\eta = \sqrt{3}\pi / 16 \approx 0.34009$. It may be a good candidate for a form of jamming point, drawing the analogy to my 2D observation for the honeycomb lattice, see Section~\ref{sub:ground}. In view of the work of Singh~\textit{et al.}~\cite{singh2020shear}, it may be a candidate geometric ground state for particles with both sliding and rolling friction. Although, the proposed value based on this argument underestimates the jamming point. It may be that this is because the analogy is not applicable, or that there is slippage in the simulations. The connection, if any, should be investigated more closely.

There are several other possibilities for generating uniform Voronoi cells. Among these, the use of a tetrahedral-octahedral honeycomb has the greatest potential. In this arrangement, the particles are the inscribed spheres to the octahedra comprising the honeycomb and each of the two tetrahedra to an octahedron contributes 1/4 of its volume to the Voronoi cell. This gives the volume fraction $\eta = 8\pi/(27\sqrt{3}) \approx 0.537422$, which is clearly not close to the glass transition in 3D. However, as also indicated in the main text, it is close to where the random-loose packing value for a frictional hard-sphere system is placed in the literature. In contrast, the honeycomb that results from the Bilinski dodecahedron has an approximate volume fraction of $\eta \approx 0.247$, the one that follows from the ten-of-diamonds decahedron has $\eta \approx 0.393$, and the one from the acute golden rhombohedron has $\eta \approx 0.424$. None is close to a known instance of dynamical arrest or jamming, and other polyhedral space fillers give similarly low volume fractions. Certainly, with the exception of the RLP value, the proposed relation is less convincing than it is in 2D.
\end{document}